\documentclass[twocolumn,aps,prl,amsmath,amssymb]{revtex4-2}

\usepackage{graphicx,bm,amsmath,amssymb,natbib,xcolor}
\usepackage{changes}

\begin{document}

\title{Electric Control of Polarity in Spin-Orbit Josephson Diode} 

\author{Junghyun Shin\textsuperscript{1}, Jae-Ho Han\textsuperscript{2,3}, Anjali Rathore\textsuperscript{4}, Joon Sue Lee\textsuperscript{4}, Seung-Bo Shim\textsuperscript{5}, Jinwoong Cha\textsuperscript{5}, Sunghun Park\textsuperscript{3*}, Junho Suh\textsuperscript{1}}
\email[Corresponding author~: ]{sunghun.park@ibs.re.kr, junhosuh@postech.ac.kr}
\affiliation{\textsuperscript{1} Department of Physics, Pohang University of Science and Technology (POSTECH), Pohang 37673, South Korea \\
\textsuperscript{2} Department of Physics, Korea Advanced Institute of Science and Technology (KAIST),Daejeon 34141, South Korea \\
\textsuperscript{3}Center for Theoretical Physics of Complex Systems, Institute of Basic Science (IBS), Daejeon 34126, South Korea\\
\textsuperscript{4}Department of Physics and Astronomy, University of Tennessee, Tennessee 37996, USA\\
\textsuperscript{5}Quantum Technology Institute, Korea Research Institute of Standards and Science (KRISS), Daejeon 34113, South Korea}
 

\begin{abstract}
The effect of spin-orbit coupling in a Josephson diode has not been elucidated due to its interplay with the complexity of Josephson devices. Here, we systematically control local electric fields in epitaxial Al-InAs Josephson junctions under in-plane magnetic fields and observe a polarity reversal of the Josephson diode. We interpret this polarity reversal as an effect of field-tunable spin-orbit coupling on nonreciprocal Josephson currents. A theoretical model, accounting for Rashba and Dresselhaus spin-orbit couplings in a planar Josephson junction containing many transverse subbands, aligns with the observed polarity reversal and its dependence on magnetic field. Our finding addresses spin-orbit control in a Josephson diode, enabling manipulation of Josephson harmonics. 
\end{abstract}
\maketitle 

Josephson junctions (JJs) are essential elements of superconducting circuits, attracting considerable experimental and theoretical interest for quantum devices \cite{Josephson1974,Makhlin2001,Martinis2020,Blais2020,Clerk2020,Blais2021}. The Josephson current flowing through a weak link depends on the material properties, symmetries, and geometry of the weak link \cite{tinkham1996,RevModPhys.76.411,RevModPhys.51.101,Kulik1970,Beenakker1991,Furusaki2002,Bagwell1992}. Recently, there has been growing attention to the nonreciprocal Josephson current, known as Josephson diode effect (JDE) \cite{wu2022, baumgartner2022,pal2022,jeon2022, matsuo2023,trahms2023,le2024}, by breaking time-reversal and inversion symmetries, which is being pursued for dissipationless superconducting electronics and rectifiers \cite{nadeem2023,ando2020}. 

The JDE is manifested in the current-phase relation (CPR) of a JJ, the minimal form of which includes the second harmonic \cite{baumgartner2022, costa2023e},    
\begin{equation}\label{CPR}
	I(\varphi) = a_{1}\sin{\left(\varphi+\varphi_1\right)}+a_{2}\sin{\left(2\varphi+\varphi_2\right)},    
\end{equation}
where $\varphi$ is the Josephson phase. $a_1$ and $a_2$ are the amplitudes, and $\varphi_1$ and $\varphi_2$ are the phase offsets of the first and second harmonics, respectively. Assuming $|a_2/a_1|\ll 1$, the Josephson diode efficiency $\eta_J$ is \cite{pal2022},
\begin{equation}\label{eta_J}
  \eta_J=\frac{I^{+}_c-I^{-}_c}{I^{+}_c+I^{-}_c} = -\frac{a_2}{a_1}\sin \delta,  
\end{equation}
where a key parameter for the onset of the diode effect is the anomalous phase difference $\delta=\varphi_2-2 \varphi_1$, which deviates from $0$ and $\pi$. Here, $I^{+}_c$ and $I^{-}_c$ are the forward and backward critical currents, respectively. Theoretically, such a CPR has been predicted for a finite Cooper-pair momentum (fCPM), as seen in the Fulde-Ferrell state \cite{Fulde1964}, helical superconducting states \cite{yuan2022}, and nanowire-superconductor hybrid structures \cite{davydova2022}, or resulting from an interplay of spin-orbit coupling (SOC) and Zeeman field in planar geometry JJs with several transverse subbands \cite{yokoyama2014}. It can also arise from supercurrent interference in a superconducting quantum interference device (SQUID) with asymmetric JJs \cite{Souto2022}. The first two mechanisms, to some extent, explain the experimental observations in Al-InAs JJs in planar \cite{baumgartner2022,banerjee2023,lotfizadeh2024} or SQUID \cite{reinhardt2024} geometries, and in a Nb-NiTe$_2$-Nb JJ \cite{pal2022}, while the last one in sufficiently asymmetric SQUID arms \cite{ciaccia2023}. 

The ability to control and understand the underlying principle of the Josephson diode polarity, the sign of $\eta_J$ in Eq.~\eqref{eta_J}, is important for device applications. The reversal of the polarity controlled by an in-plane magnetic field has been observed, though not exclusively, in planar Al-InAs JJs, which can be linked to $0$-$\pi$ transitions in Josephson inductance measurements \cite{costa2023e}, a topological phase transition in tunneling spectroscopy \cite{banerjee2023}, or the physical lengths of electrodes in transport measurements \cite{lotfizadeh2024}. 
While they suggest that the occurrence of the polarity reversal is responsible for intricate effects, including not only SOC and Zeeman field but also device geometry, control experiments to identify the underlying principle have been elusive. 

In this work, we investigate supercurrent transports in a Josephson device to elucidate the SOC effect on JDE. The device is made of two planar JJs forming a SQUID where each JJ is covered by a top gate electrode, allowing for electric field control in the individual junctions. We focus on the JDE and CPR, revealed by SQUID oscillations, for the identical JJs, set by tuning the gate voltages symmetrically, measuring the diode efficiency and anomalous phase difference as a function of an in-plane magnetic field $B_y$. A distinguishing feature is their strong dependence on the gate voltage at high magnetic fields $B_y>30$ mT, showing a polarity reversal, accompanying a $\pi$-crossing of the anomalous phase difference, without signatures of a 0-$\pi$ transition \cite{costa2023e} or a topological phase transition \cite{banerjee2023}. 
Our theoretical model attributes the polarity reversal to the coherent interplay between anisotropic SOC, including Rashba and Dresselhaus terms, and fCPM from the orbital effect of $B_y$. The interplay effect is sensitive to the ratio between the Rashba and Dresselhaus SOC, which is tunable with the gate voltage, offering SOC-driven electrical control of CPR harmonics in Josephson junctions.

\begin{figure}[t!]%
\centering
\includegraphics[width=0.9 \columnwidth]{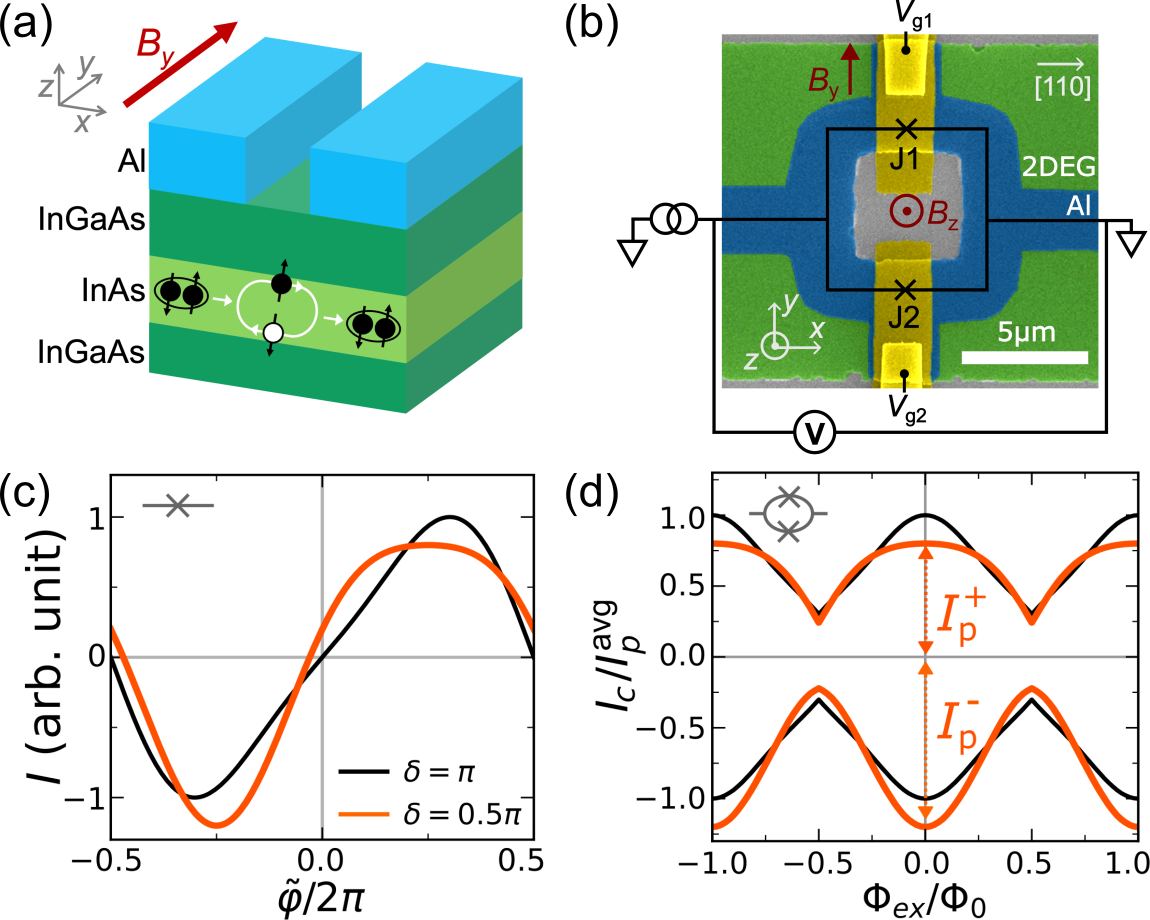}
\caption{
(a) Schematic cross-section of a planar JJ in a hybrid Al-InAs heterostructure. The InAs quantum well is topped by a 10-nm-thick InGaAs barrier, covered by a 6-nm-thick Al layer. 
(b) False-colored scanning electron microscope image of a SQUID fabricated on an Al-InAs two-dimensional heterostructure. Two planar JJs, denoted by J1 and J2, are controlled by local electric fields via top gate voltages, $V_{g1}$ and $V_{g2}$. An in-plane magnetic field $B_y$ is applied perpendicular to DC bias current which is parallel to the [110] crystallographic direction of the InAs layer indicated by the white arrow. An out-of-plane magnetic field $B_z$ threads the external flux $\Phi_{ex}$ through the SQUID loop.
(c) The CPR, $I(\tilde{\varphi})$, of a single JJ given in Eq.~\eqref{CPR}, where $\tilde{\varphi}=\varphi+\varphi_1$. The black (orange) curve represents $\delta=\pi$ ($\delta=0.5\pi$) with $a_2/a_1=0.2$. These parameters are chosen for illustration. 
(d) $I_{c}$ oscillations as a function of $\Phi_{ex}/\Phi_0$ for a symmetric SQUID, where $\Phi_{0}$ is the flux quantum $h/2e$. Each junction of the SQUID follows the CPR shown in (c). The color codes indicate the same parameters as in (c). Oscillations are normalized by $I_p^{\text{avg}} = (I_p^+ + I_p^-)/2$.
}
\label{fig1}
\end{figure}

Our device is a DC SQUID comprised of two planar superconductor-normal-superconductor JJs fabricated with an epitaxial Al-InAs heterostructure shown in Figs.~\ref{fig1}(a) and \ref{fig1}(b). The JJs are realized by removing strips of the Al layer \cite{SM}\nocite{shabani2016,kjaergaard2016,knap1996,lin2022,brandt2004,mayer2020,clarke2004,Gennes1989,Scharf2019,lee2019,Clem2010}. Both junctions have the nominal junction length of $L_{j}=100$ nm and width of $W=4.5$ $\mu$m. A large $B_y$ up to 200 mT is employed to break time-reversal symmetry, while sub-mT $B_z$ is used to adjust the external flux $\Phi_{\text{ex}}$ passing through the SQUID loop. 

The anomalous phase difference $\delta$ in Eq.~\eqref{eta_J} determines the JDE polarity (Fig.~\ref{fig1}(c)). Assuming $0 \leq \delta < 2\pi$, when $0 < \delta < \pi$, $I_{c}^{+} < I_{c}^{-}$, and vice versa for \(\pi<\delta<2\pi\). In a SQUID, the shape of $I_c$ oscillations is governed by the CPRs of the JJs. Figure~\ref{fig1}(d) shows representative SQUID oscillations in a symmetric SQUID, whose CPRs correspond to those in Fig.~\ref{fig1}(c). The forward and backward peaks, $I_{p}^{+}$ and $I_{p}^{-}$, correspond to the maximum constructive interference, which is the sum of the critical currents from each JJ in that direction. We quantify the nonreciprocity as $\eta = (I_{p}^{+} - I_{p}^{-})/(I_{p}^{+} + I_{p}^{-})$ which equals the Josephson diode efficiency of identical JJs, $\eta=\eta_J$. This nonreciprocity is distinct from that in asymmetric SQUIDs \cite{Souto2022,ciaccia2023, SM}.

\begin{figure}[t!]%
\centering
\includegraphics[width= 0.9 \columnwidth]{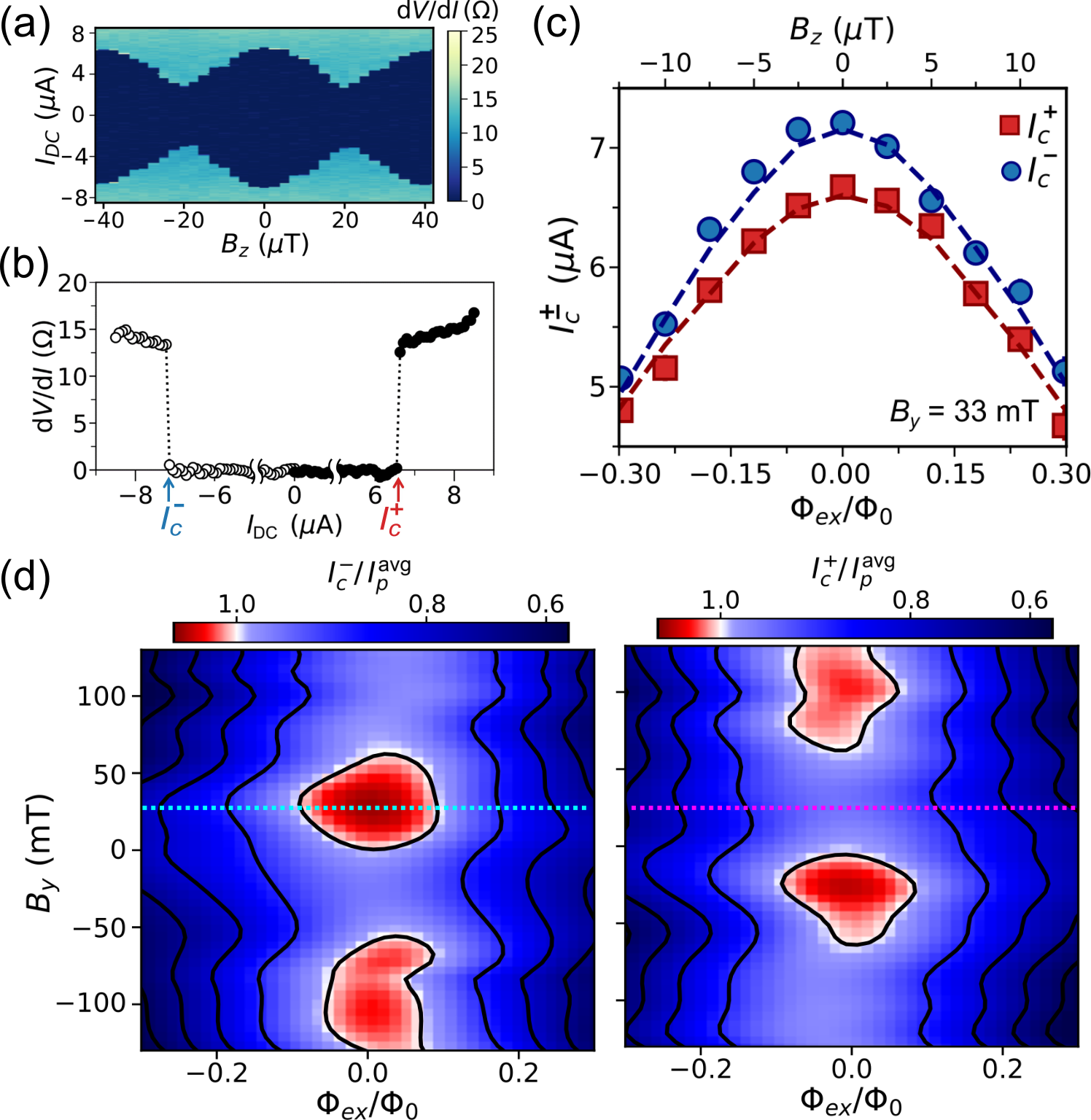}
\caption{
(a) $dV/dI$ versus $I_{\text{DC}}$ and $B_z$ at $V_g = 0$ and $B_y = 33$ mT.  
(b) $dV/dI$ as a function of $I_{\text{DC}}$ at $B_y = 33$ mT, $B_z = 0$, and $V_g = 0$, with arrows indicating $I_c^+$ and $I_c^-$.  
(c) $I_c^+$ (red squares) and $I_c^-$ (blue circles) near the SQUID oscillation peak at $V_g = 0$, $B_y = 33$ mT. Dashed lines are numerical fits \cite{SM}.  
(d) 2D maps of $I_c^-$ (left) and $I_c^+$ (right) versus $\Phi_{\text{ex}}/\Phi_0$ and $B_y$, normalized by $I_p^{\text{avg}}$ for each $B_y$. 
Black contours range from 0.6 to 1 in steps of 0.1. The dotted line marks $B_y = 33$ mT, as shown in (c).
}
\label{fig2}
\end{figure}

We measure differential resistance (d$V$/d$I$) as a function of $B_z$ and $I_{DC}$ at $B_y=33$ mT and $V_{g} = 0$ (Figs.~\ref{fig2}(a) and \ref{fig2}(b)). While the critical currents oscillate with a fixed period (Fig.~\ref{fig2}(a)), their magnitude depends on the current direction (Fig.~\ref{fig2}(b)). The difference between $I_{c}^{+}$ and $ I_{c}^{-}$ is most pronounced near the peaks (Fig.~\ref{fig2}(c)). This is attributed to the JDE in the JJs, resulting in unidirectional supercurrents when the current amplitude falls within the range between $I_{c}^{+}$ and $I_{c}^{-}$ \cite{SM}.

Figure~\ref{fig2}(d) shows $I_{c}^{+}$ and $I_{c}^{-}$ near $\Phi_{\text{ex}}= 0 $ at various $B_{y}$, normalized by the average peak $I_{p}^{\text{avg}}$ for each $B_{y}$. Without JDE, both $I_{c}^{+}$ and $I_{c}^{-}$ should equal $I_{p}^{\text{avg}}$ at $\Phi_{ex} = 0$. However, we observe that $I_{c}^{+}$ and  $I_{c}^{-}$ at $\Phi_{\text{ex}} = 0$ show modulation with respect to $B_{y}$ in opposite directions. In addition, the JDE polarity changes over $B_{y}$, reversing at $B_{y}=0$ and $B_{y}\approx\pm 55$ mT. We note that the polarity reversal at finite magnetic field implies nontrivial mechanisms governing JDE in our device as the field strength is significantly lower than that observed in previous experiments, within the 220--400 mT range \cite{costa2023e, banerjee2023, lotfizadeh2024}.

\begin{figure}[t!]%
\centering
\includegraphics[width=0.95\columnwidth]{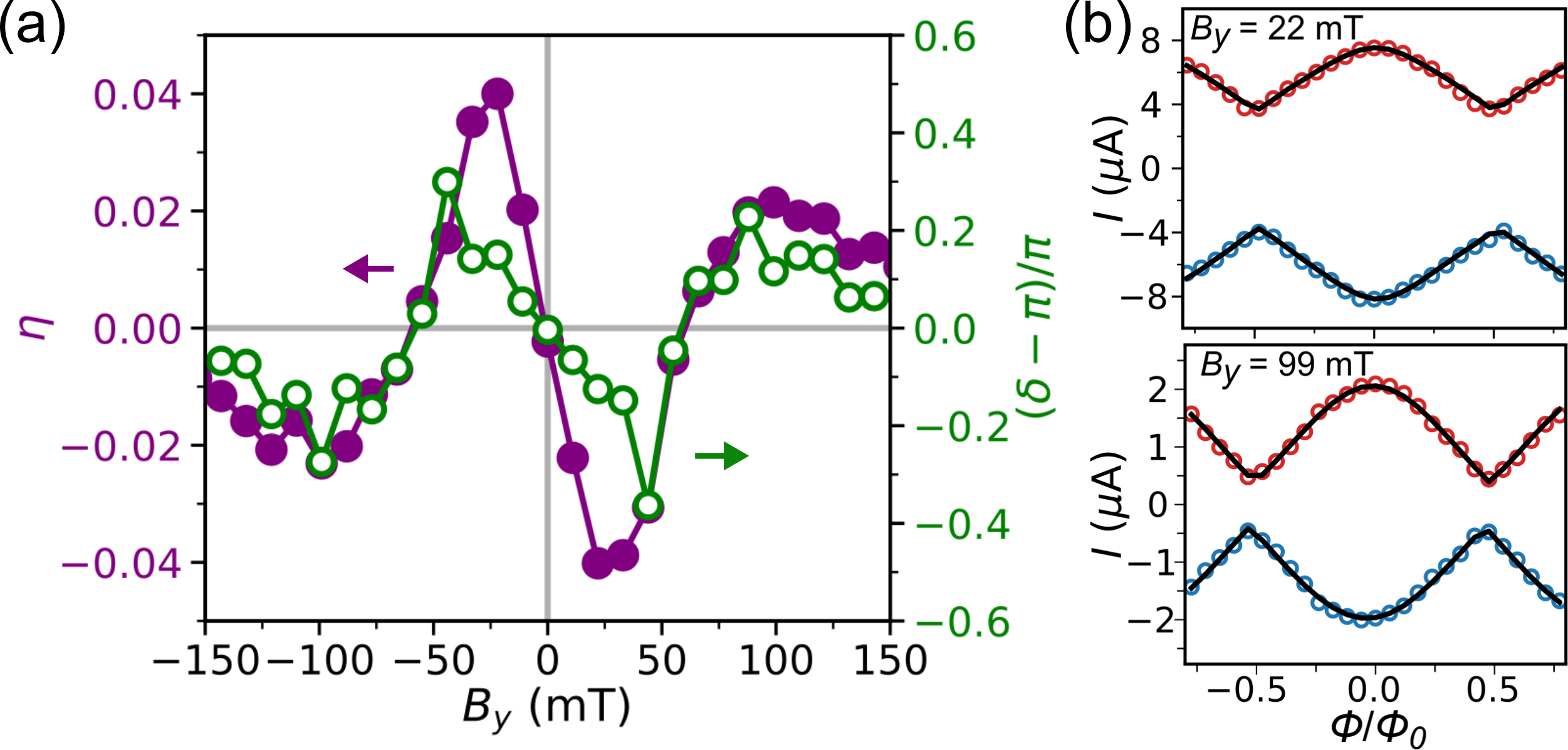}
\caption{
Extracted diode efficiency $\eta$ and anomalous phase shift $\delta$.
(a) $\eta$ and $\delta$ versus $B_y$ at $V_g = 0$.  
(b) SQUID oscillations of $I_c^+$ and $I_c^-$ at $B_y$ = 22 and 99 mT, with fits from Eq.~(1) \cite{SM}.
}
\label{fig3}
\end{figure}

Figure~\ref{fig3}(a) presents the nonreciprocal efficiency $\eta$ as a function of $B_{y}$, which is obtained from the data shown in Fig.~\ref{fig2}(d). $\eta$ is anti-symmetric with respect to $B_y$, consistent with the Onsager-Casimir relations \cite{Casimir1945,banerjee2023}. In addition, the diode efficiency displays strong dependence on $B_{y}$ in a non-monotonic way. Notably, it increases linearly from $B_{y}$ = 0, decreases beyond 33 mT, and reverses sign beyond 55 mT. This polarity-reversed diode efficiency peaks around $B_{y}$ = 100 mT and gradually diminishes with further increase in $B_{y}$. The magnitude of $\eta$ reaches maximum at approximately 0.04, reaching 0.02 after polarity reversal.

We extract $\delta$ by fitting the measured SQUID oscillations, as shown in Fig.~\ref{fig3}(b), using current conservation and flux quantization (see Ref. \cite{SM} for details). The extracted $\delta$, plotted in Fig.~\ref{fig3}(a), shows a direct correlation with $\eta$. When $0<\delta<\pi$, $\eta<0$, and when $\pi<\delta<2\pi$, $\eta>0$, thus the polarity reversal coincides with $\delta=\pi$. This correspondence between $\delta$ and $\eta$ implies the existence of higher harmonics in CPR (Eq.~\eqref{CPR}).

\begin{figure}[t!]%
\centering
\includegraphics[width=0.95\columnwidth]{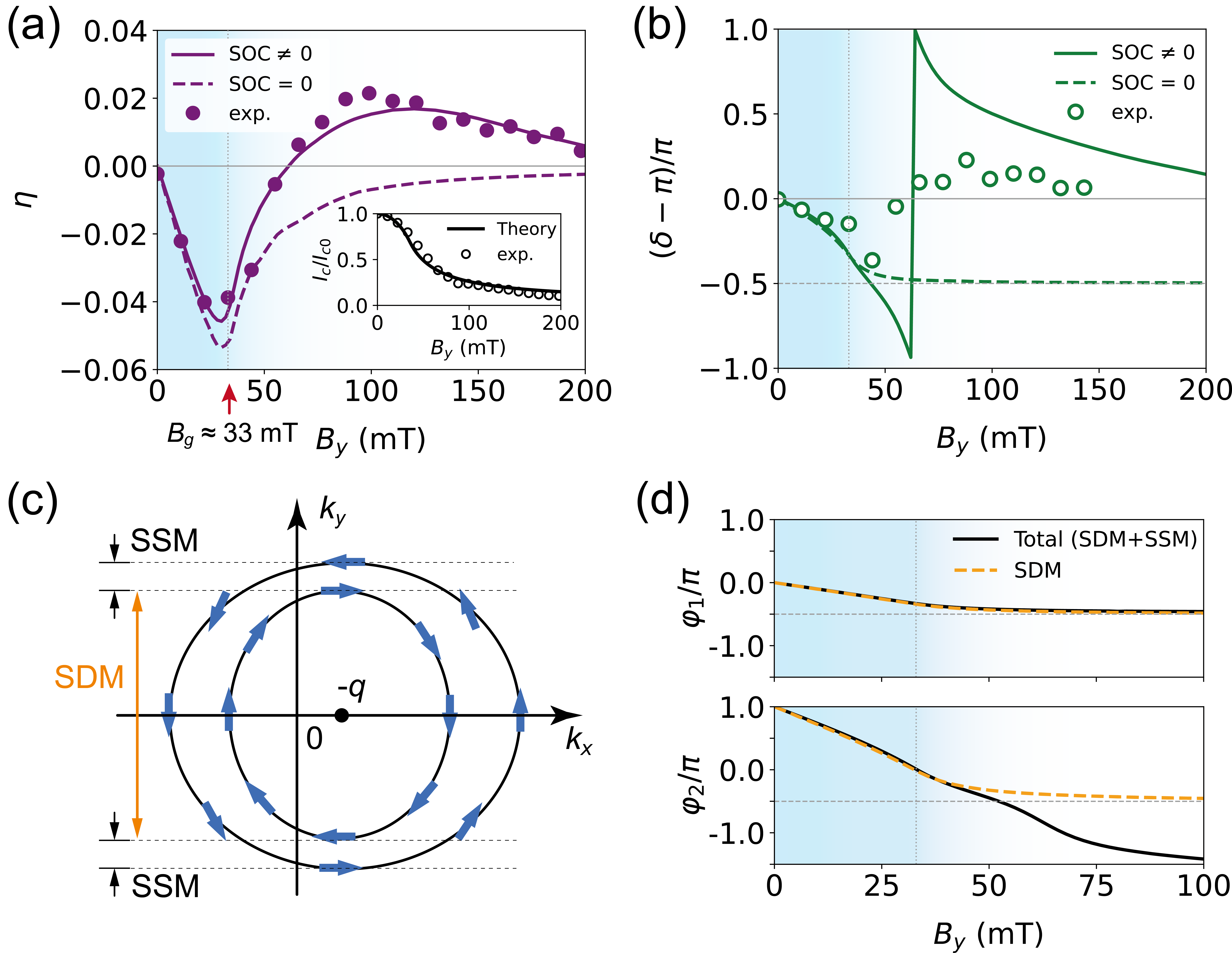}
\caption{
Calculated diode efficiency $\eta$ and anomalous phase shift $\delta$.
(a) $\eta$ with and without SOC. The efficiency below $B_g\approx 33$ mT is fCPM-dominated (blue-shaded), and gradually changes to the SOC-assistant regime (white) as the $B_y$ increases. That above $B_g$ (white) is affected by the SOC. Inset: $B_y$ dependence of normalized critical current.
(b) $\delta$ with and without SOC, obtained by fitting the numerical results in (a) with the minimal CPR model in Eq.~\eqref{CPR}. In (a) and (b), the data in Fig.~\ref{fig3}(a) are presented together for comparison. 
(c) Schematic of the Fermi surfaces with SOC and fCPM. The center shifts by the momentum, $-q$, and the Fermi surfaces are split due to the SOC. The blue arrows denote spin direction with the momentum dependence induced by the SOC. 
(d) The first and the second anomalous phase shifts, $\varphi_1$ and $\varphi_2$. The phase shifts, illustrated by the black solid lines, correspond to the case of $q \neq 0$ and SOC $\neq$ 0 shown in (a) and (b). The orange dashed lines depict the phase shifts from SDM only.
}
\label{fig4}
\end{figure}

In our device, SOC and fCPM are responsible for the nonreciprocal critical currents. Rashba ($\alpha$) and Dresselhaus ($\beta$) SOCs arise from inversion symmetry breaking by electric fields, $\alpha$ from an external field and $\beta$ from an intrinsic field within the crystal lattice. The current flowing along the [110] direction of the InAs layer leads to an anisotropic SOC owing to the coexistence of Rashba and Dresselhaus SOCs \cite{Baumgartner_2022}. Additionally, when applying $B_y$, Cooper pairs acquire fCPM due to the orbital effect \cite{davydova2022}. In the vector potential $\textbf{A}=B_y z \hat{x}$, parallel to the interfaces of the superconductor ($z=d$) and the semiconductor ($z=0$), the orbital effect leads to the fCPM, $q=-\pi B_y d/\Phi_0$. The fCPM decreases and eventually closes the proximity-induced gap at $B_y = B_g = \Delta/(e v_F d)$ due to the Doppler shift of Bogoliubov quasiparticle energy $\Delta\pm \hbar v_F q$ \cite{davydova2022}. While the Zeeman effect with SOC \cite{yuan2022,daido2022, Legg2022} also leads to fCPM, it is estimated to be three orders of magnitude smaller than the orbital effect and thus is neglected in our model \cite{SM}.

We use a theoretical model of a short Josephson junction that accounts for the interplay between SOC and fCPM and fit the $B_{y}$ dependence of the efficiency $\eta$ by varying the chemical potential $\mu$, the junction transparency $\tau$ and the Rashba SOC $\alpha$, with constant Dresselhaus SOC $\beta$ (Fig.~\ref{fig4}(a)). Details of fitting parameters are summarized in Supplemental Material \cite{SM}. From the best fit, we estimate $B_g \approx 33$ mT, with induced superconducting gap $\Delta =170$ $\mu$eV and $v_F = 5.1\times 10^5 $ m$\cdot$s$^{-1}$. The relative importance of SOC and fCPM becomes evident when comparing our model to an alternative model without SOC. In the low-field region of $B_y < B_g$, $\eta$ from the SOC-free model also follows the data closely, indicating that the fCPM dominates JDE at this region. The $B_{y}$ dependence of $I_{p}^{\text{avg}}$ supports the crucial role of the fCPM. $I_{p}^{\text{avg}}$ gradually decreases with increasing $B_{y}$, becoming roughly half near 55 mT (inset of Fig.~\ref{fig4}(a)). This behavior is reproduced in the model when fCPM is included, independent of SOC. In contrast, the SOC-free model does not explain the polarity reversal at high fields, demonstrating the crucial contribution of SOC to JDE. This is further evidenced by the anomalous phase difference $\delta$. In Fig.~\ref{fig4}(b), $\delta$ from the SOC-assisted and SOC-free models diverge significantly at $B_{y} > B_{g}$. $\delta$ saturates at $\pi/2$ in the SOC-free model, whereas the SOC-assisted model predicts that $\delta$ does not saturate at $\pi/2$ and eventually reaches the opposite-polarity region, transitioning from $ 0 < \delta < \pi$ to $\pi <\delta < 2\pi$.

To further examine the SOC effect on JDE, we analyze two types of transverse modes, spin-degenerate modes (SDM) and spin-split modes (SSM) on the Fermi surfaces with spin-orbit splitting (Fig.~\ref{fig4}(c)). Importantly, SSM is eminent in the polarity reversal observed in JDE (Fig.~\ref{fig4}(d)). The phase shifts induced by SDM reproduce those in the SOC-free model. Both $\varphi_1$ and $\varphi_2$ in Eq.~\eqref{CPR} approach $-\pi/2$ with increasing $B_{y}$, leading to the saturation of $\delta = \pi/2$ and the absence of the polarity reversal. In contrast, the contribution of SSM advances $\delta$ further to cross $\pi$, due to the spin-orbit effect on $\varphi_2$, resulting in the polarity reversal.

The analysis of SDM and SSM reveals that the emergence of the polarity reversal depends on the anisotropy of the SOC. The Andreev spectra of the SDM and SSM evolve differently with $B_y$. While the SDM shows Zeeman-like level splitting, the SSM exhibits a phase shift relative to SDM. The distinct evolution produces higher harmonics in the total CPR. For Rashba-only SOC, this effect is weak as the contribution from SSM is only a small fraction of the total channels, on the order of $k_{\text{SOC}}/k_F \ll 1$, and does not significantly influence the higher harmonics. Here, $k_{\text{SOC}}=2m^* \alpha/\hbar^2$ and $k_F$ are the SOC-induced wavevector splitting and the Fermi wavevector, respectively. In contrast, the anisotropic SOC, with comparable $\alpha$ and $\beta$, enhances the amplitude of CPR of the SSM, making it comparable to the second harmonic of SDM in the SOC-assisted regime \cite{SM}. Thus, SSM can significantly contribute to the phase shift $\varphi_2$, causing $\delta$ to cross $\pi$ to result in the reversal of JDE polarity.

\begin{figure}[t!]%
\centering
\includegraphics[width=0.9\columnwidth]{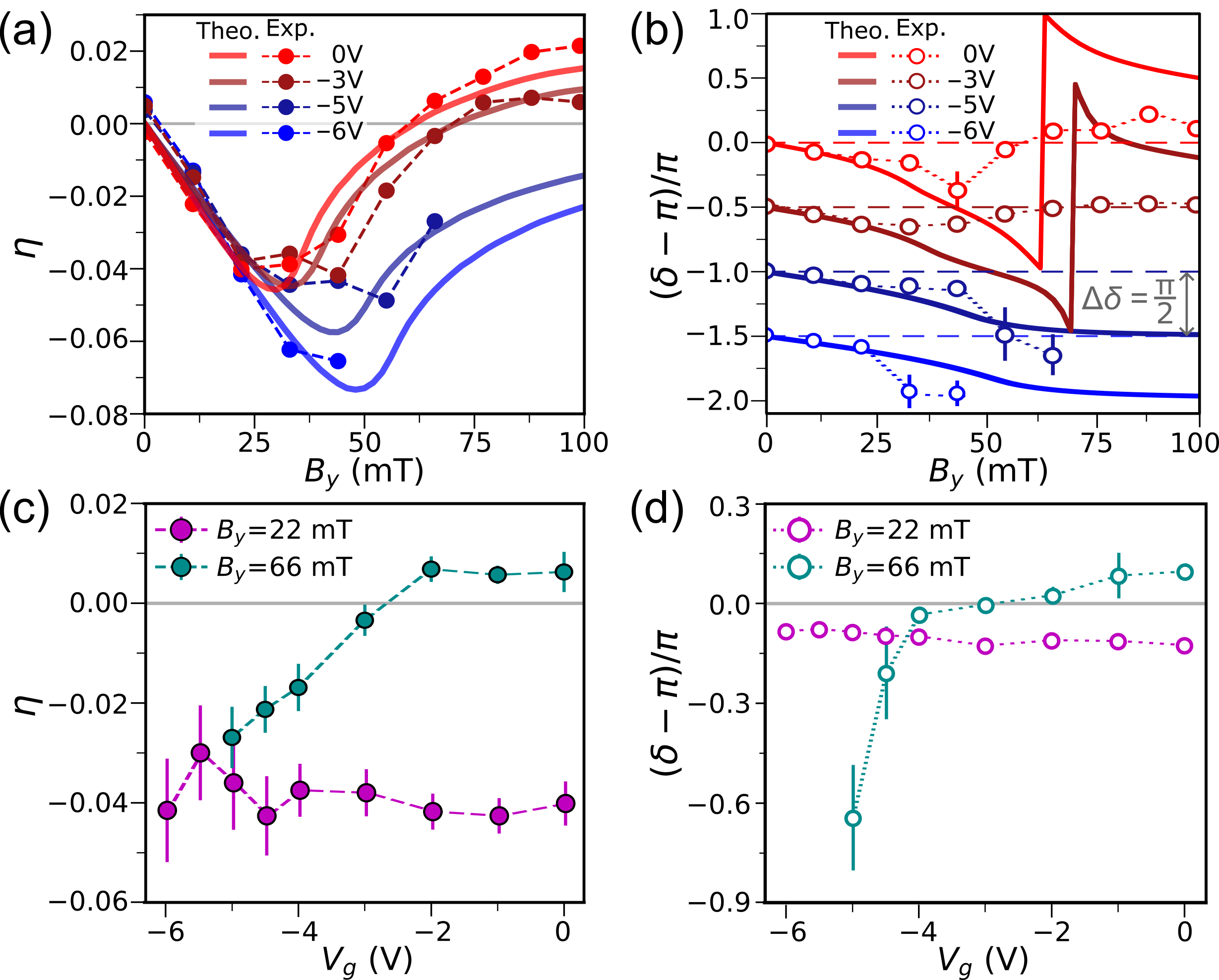}
\caption{
Electric control of the JDE polarity.
(a) $B_y$ dependence of the diode efficiency ($\eta$) for $V_g$ from 0 V to $-6$ V, with fits from our model (parameters in Table S1, Supplemental Material \cite{SM}).  
(b) Anomalous phase difference ($\delta$) versus $B_y$ for different $V_g$. Solid lines are theoretical results obtained by fitting the numerical results from (a). Data are vertically offset for clarity.  
(c), (d) The evolution of $\eta$ (c) and $\delta$ (d) with $V_g$ at $B_y$ = 22 and 66 mT.
}
\label{fig5}
\end{figure}

We attribute the JDE polarity reversal to the coherent interplay between the fCPM and the anisotropic SOC. The Rashba SOC strength is controllable via gate voltage \cite{SM}, enabling tuning of the SOC characteristics in the InAs layer. By adjusting $V_{g}$ from 0 V to $-6$ V, we obtained the $B_y$ dependence of $\eta$ and $\delta$ (Figs.~\ref{fig5}(a) and \ref{fig5}(b)), showing the vanishing of high-field sign-reversal, along with the evolution from the $\pi$-crossing behavior to the $\pi$/2-saturation. Gate voltage leads to distinct effects in the fCPM-dominated regime at low fields and the SOC-assisted regime at high fields. The fCPM-dominated regime at $B_y < B_g$ is nearly independent of $V_g$, while the SOC-assisted regime at $B_y > B_g$ is highly adjustable by $V_g$. Figures~\ref{fig5}(c) and \ref{fig5}(d) illustrate the $V_g$ dependence at two selected fields, 22 mT and 66 mT, representing the two distinct regimes. At $B_y = 22$ mT, $\delta$ is weakly dependent on $V_g$, with $\eta \approx -0.04$. Conversely, at $B_y = 66$ mT, $\delta$ is strongly affected by $V_g$, exhibiting $\pi$-crossing at $V_g \approx -3$ V, with $\eta$ varying from approximately $-0.03$ to $0.01$.

To assess the gate voltage effect on our model calculations, we gradually decrease the chemical potential as $V_g$ decreases and adjust $\tau$ and $\alpha$, respectively. The resulting fits in $\eta$ exhibit excellent agreement with the experimental data at both zero and finite $V_g$ (Fig.~\ref{fig5}(a)). The evolution of $\delta$ reproduces the transition from the $\pi$-crossing behavior to the $\pi$/2-saturation (Fig.~\ref{fig5}(b)). This confirms that the observed polarity reversal, which is controllable using the gate voltage or the magnetic field, results from the coherent interplay between SOC and fCPM.

Previously reported polarity reversal in Al-InAs JJs was explained by a 0-$\pi$ transition driven by the Zeeman energy \cite{costa2023e,yokoyama2014} or a topological phase transition \cite{banerjee2023}. The former mechanism arises at $|g\mu_B B_y|/2 \approx \hbar v_F/L_j$ from which we estimate $B_y \approx 6.9$ T, much higher than the field strength in our experiments \cite{costa2023e}. In addition, the $B_y$ dependence of the critical current of J2, while J1 is almost pinched off \cite{SM} does not show the signature of the topological phase transition --- the suppression and revival of the critical current \cite{Dartiailh2021} --- ruling out the possibility of topological transition in our experiments. 

Our model shows how a combination of Rashba and Dresselhaus SOCs can induce the polarity reversal for a range of experimental parameters. Analyzing the phase diagram of JDE in $V_{g}$--$B_{y}$ space allows us to find the optimal parameter region for controlling the polarity reversal using the gate voltage. Our findings introduce an additional tuning parameter for controlling JDE.  Although we focus on the [110] crystallographic direction, further studies are needed to examine JDE along other crystallographic axes, where SOC variations may yield different gate-voltage dependencies.

In conclusion, we demonstrate electric control of the JDE with high tunability to the extent that its polarity is reversed. This controllability of JDE is provided by adjusting gate voltages under in-plane magnetic field perpendicular to the supercurrents. Our theoretical model shows that the higher harmonics of CPR determine this polarity reversal. Analysis of multichannel contributions to the CPR reveals that the gate-voltage induced changes of spin-orbit anisotropy perturb the higher harmonics, affecting the JDE polarity as a result. At the point of polarity reversal where JDE becomes zero, the device behaves as if Rashba SOC effectively recovers the broken symmetries. Our device is compatible with superconducting quantum circuit architectures \cite{Casparis2018} and its locally tunable nonreciprocity could lead to novel applications in superconducting electronic devices \cite{monroe2022,monroe2024}.

\section*{Acknowledgments}
J. Suh acknowledges supports from Samsung Foundation(SSTF-BA1801-03) and National Research Foundation(RS-2023-00207732, RS-2024-00352688, 2022M3H3A1064154). S.P. acknowledges the support from the Institute for Basic Science (IBS) in the Republic of Korea through the project IBS-R024-Y4. 

\bibliography{JDE}

\begin{thebibliography}{10}

\bibitem{Josephson1974}
B.~D. Josephson, {\it The discovery of tunnelling supercurrents\/}, Rev. Mod.
  Phys. {\bf 46}, 251 (1974).

\bibitem{Makhlin2001}
Y.~Makhlin, G.~Sch\"on, and A.~Shnirman, {\it Quantum-state engineering with
  {J}osephson-junction devices\/}, Rev. Mod. Phys. {\bf 73}, 357 (2001).

\bibitem{Martinis2020}
J.~M. Martinis, M.~H. Devoret, and J.~Clarke, {\it Quantum {J}osephson junction
  circuits and the dawn of artificial atoms\/}, Nat. Phys. {\bf 16}, 234
  (2020).

\bibitem{Blais2020}
A.~Blais, S.~M. Girvin, and W.~D. Oliver, {\it Quantum information processing
  and quantum optics with circuit quantum electrodynamics\/}, Nat. Phys. {\bf
  16}, 247 (2020).

\bibitem{Clerk2020}
A.~A. Clerk, K.~W. Lehnert P.~Bertet, J.~R. Petta, and Y.~Nakamura, {\it Hybrid
  quantum systems with circuit quantum electrodynamics\/}, Nat. Phys. {\bf 16},
  257 (2020).

\bibitem{Blais2021}
A.~Blais A.~L. Grimsmo, S.~M. Girvin, and A.~Wallraff, {\it Circuit quantum
  electrodynamics\/}, Rev. Mod. Phys. {\bf 93}, 025005 (2021).

\bibitem{tinkham1996}
M.~Tinkham, {\it Introduction to Superconductivity\/} (McGraw-Hill, 1996),
  second edn.

\bibitem{RevModPhys.76.411}
A.~A. Golubov, M.~Y. Kupriyanov, and E.~Il'ichev, {\it The current-phase
  relation in Josephson junctions\/}, Rev. Mod. Phys. {\bf 76}, 411 (2004).

\bibitem{RevModPhys.51.101}
K.~K. Likharev, {\it Superconducting weak links\/}, Rev. Mod. Phys. {\bf 51},
  101 (1979).

\bibitem{Kulik1970}
I.~O. Kulik, {\it Macroscopic Quantization and the Proximity Effect in S-N-S
  {J}unctions\/}, JETP {\bf 30}, 944 (1970).

\bibitem{Beenakker1991}
C.~W.~J. Beenakker, {\it Universal limit of critical-current fluctuations in
  mesoscopic {J}osephson junctions\/}, Phys. Rev. Lett. {\bf 67}, 3836 (1991).

\bibitem{Furusaki2002}
A.~Furusaki and M.~Tsukada, {\it Dc Josephson effect and Andreev reflection\/},
  Solid State Communications {\bf 78}, 299 (1991).

\bibitem{Bagwell1992}
P.~F. Bagwell, {\it Suppression of the Josephson current through a narrow,
  mesoscopic, semiconductor channel by a single impurity\/}, Phys. Rev. B {\bf
  46}, 12573 (1992).

\bibitem{wu2022}
H.~Wu, Y.~Wang, Y.~Xu, P.~K. Sivakumar, C.~Pasco, U.~Filippozzi, S.~S.~P.
  Parkin Y.-J. Zeng, T.~{McQueen}, and M.~N. Ali, {\it The field-free
  {J}osephson diode in a van der {W}aals heterostructure\/}, Nature {\bf 604},
  653 (2022).

\bibitem{baumgartner2022}
C.~Baumgartner, L.~Fuchs, A.~Costa, S.~Reinhardt, S.~Gronin, G.~C. Gardner,
  T.~Lindemann, M.~J. Manfra, P.~E. Faria~Junior, D.~Kochan J.~Fabian,
  N.~Paradiso, and C.~Strunk, {\it Supercurrent rectification and magnetochiral
  effects in symmetric {J}osephson junctions\/}, Nat. Nanotechnol. {\bf 17}, 39
  (2022).

\bibitem{pal2022}
B.~Pal, A.~Chakraborty, P.~K. Sivakumar, M.~Davydova, A.~K. Gopi, A.~K.
  Pandeya, J.~A. Krieger, Y.~Zhang, M.~Date, S.~Ju, N.~Yuan N.~B.~M. Schröter,
  L.~Fu, and S.~S.~P. Parkin, {\it Josephson diode effect from {C}ooper pair
  momentum in a topological semimetal\/}, Nat. Phys. {\bf 18}, 1228 (2022).

\bibitem{jeon2022}
K.-R. Jeon, J.-K. Kim, J.~Yoon, J.-C. Jeon, H.~Han A.~Cottet, T.~Kontos, and
  S.~S.~P. Parkin, {\it Zero-field polarity-reversible {J}osephson supercurrent
  diodes enabled by a proximity-magnetized {Pt} barrier\/}, Nat. Mater. {\bf
  21}, 1008 (2022).

\bibitem{matsuo2023}
S.~Matsuo, T.~Imoto, T.~Yokoyama, Y.~Sato, T.~Lindemann, S.~Gronin G.~C.
  Gardner, M.~J. Manfra, and S.~Tarucha, {\it Josephson diode effect derived
  from short-range coherent coupling\/}, Nat. Phys. {\bf 19}, 1636 (2023).

\bibitem{trahms2023}
M.~Trahms, L.~Melischek, J.~F. Steiner, B.~Mahendru, I.~Tamir, N.~Bogdanoff,
  O.~Peters, G.~Reecht C.~B. Winkelmann, F.~Von~Oppen, and K.~J. Franke, {\it
  Diode effect in Josephson junctions with a single magnetic atom\/}, Nature
  {\bf 615}, 628 (2023).

\bibitem{le2024}
T.~Le, Z.~Pan, Z.~Xu, J.~Liu, J.~Wang, Z.~Lou, X.~Yang, Z.~Wang Y.~Yao, C.~Wu,
  and X.~Lin, {\it Superconducting diode effect and interference patterns in
  kagome {CsV}3Sb5\/}, Nature {\bf 630}, 64 (2024).

\bibitem{nadeem2023}
M.~Nadeem, M.~S. Fuhrer, and X.~Wang, {\it The superconducting diode effect\/},
  Nat. Rev. Phys. {\bf 5}, 558 (2023).

\bibitem{ando2020}
F.~Ando, Y.~Miyasaka, T.~Li, J.~Ishizuka, T.~Arakawa, Y.~Shiota T.~Moriyama,
  Y.~Yanase, and T.~Ono, {\it Observation of superconducting diode effect\/},
  Nature {\bf 584}, 373 (2020).

\bibitem{costa2023e}
A.~Costa, C.~Baumgartner, S.~Reinhardt, J.~Berger, S.~Gronin, G.~C. Gardner,
  T.~Lindemann, M.~J. Manfra, J.~Fabian D.~Kochan, N.~Paradiso, and C.~Strunk,
  {\it Sign reversal of the {J}osephson inductance magnetochiral anisotropy and
  0–$\pi$-like transitions in supercurrent diodes\/}, Nat. Nanotechnol. {\bf
  18}, 1266 (2023).

\bibitem{Fulde1964}
P.~Fulde and R.~A. Ferrell, {\it Superconductivity in a Strong Spin-Exchange
  Field\/}, Phys. Rev. {\bf 135}, A550 (1964).

\bibitem{yuan2022}
N.~F.~Q. Yuan and L.~Fu, {\it Supercurrent diode effect and finite-momentum
  superconductors\/}, Proc. Natl. Acad. Sci. USA {\bf 119}, e2119548119 (2022).

\bibitem{davydova2022}
M.~Davydova, S.~Prembabu, and L.~Fu, {\it Universal {J}osephson diode
  effect\/}, Sci. Adv. {\bf 8}, eabo0309 (2022).

\bibitem{yokoyama2014}
T.~Yokoyama, M.~Eto, and Y.~V. Nazarov, {\it Anomalous {J}osephson effect
  induced by spin-orbit interaction and {Z}eeman effect in semiconductor
  nanowires\/}, Phys. Rev. B {\bf 89}, 195407 (2014).

\bibitem{Souto2022}
R.~S. Souto, M.~Leijnse, and C.~Schrade, {\it Josephson Diode Effect in
  Supercurrent Interferometers\/}, Phys. Rev. Lett. {\bf 129}, 267702 (2022).

\bibitem{banerjee2023}
A.~Banerjee, M.~Geier, M.~A. Rahman, C.~Thomas, T.~Wang M.~J. Manfra,
  K.~Flensberg, and C.~M. Marcus, {\it Phase Asymmetry of {A}ndreev Spectra
  from {C}ooper-Pair Momentum\/}, Phys. Rev. Lett. {\bf 131}, 196301 (2023).

\bibitem{lotfizadeh2024}
N.~Lotfizadeh, W.~F. Schiela, B.~Pekerten, P.~Yu, B.~H. Elfeky W.~M.
  Strickland, A.~Matos-Abiague, and J.~Shabani, {\it Superconducting diode
  effect sign change in epitaxial {Al-InAs} {J}osephson junctions\/}, Commun.
  Phys. {\bf 7}, 120 (2024).

\bibitem{reinhardt2024}
S.~Reinhardt, T.~Ascherl, A.~Costa, J.~Berger, S.~Gronin, G.~C. Gardner,
  T.~Lindemann, M.~J. Manfra, J.~Fabian D.~Kochan, C.~Strunk, and N.~Paradiso,
  {\it Link between supercurrent diode and anomalous {J}osephson effect
  revealed by gate-controlled interferometry\/}, Nat. Commun. {\bf 15}, 4413
  (2024).

\bibitem{ciaccia2023}
C.~Ciaccia, R.~Haller, A.~C.~C. Drachmann, T.~Lindemann M.~J. Manfra,
  C.~Schrade, and C.~Schönenberger, {\it Gate-tunable Josephson diode in
  proximitized {InAs} supercurrent interferometers\/}, Phys. Rev. Research {\bf
  5}, 033131 (2023).

\bibitem{SM}
 See Supplemental Material at [URL will be inserted by publisher], which
  includes Refs. [34--44] for details of device characteristics, fabrication,
  measurement, additional experimental data, fitting procedure, and theoretical
  model.

\bibitem{shabani2016}
J.~Shabani, M.~Kjaergaard, H.~J. Suominen, Y.~Kim, F.~Nichele, K.~Pakrouski,
  T.~Stankevic, R.~M. Lutchyn, P.~Krogstrup, R.~Feidenhans'l, S.~Kraemer,
  C.~Nayak M.~Troyer, C.~M. Marcus, and C.~J. Palmstrøm, {\it Two-dimensional
  epitaxial superconductor-semiconductor heterostructures: A platform for
  topological superconducting networks\/}, Phys. Rev. B {\bf 93}, 155402
  (2016).

\bibitem{kjaergaard2016}
M.~Kjaergaard, F.~Nichele, H.~J. Suominen, M.~P. Nowak, M.~Wimmer, A.~R.
  Akhmerov, J.~A. Folk, K.~Flensberg J.~Shabani, C.~J. Palmstrøm, and C.~M.
  Marcus, {\it Quantized conductance doubling and hard gap in a two-dimensional
  semiconductor–superconductor heterostructure\/}, Nat. Commun. {\bf 7},
  12841 (2016).

\bibitem{knap1996}
W.~Knap, C.~Skierbiszewski, A.~Zduniak, E.~Litwin-Staszewska, D.~Bertho,
  F.~Kobbi, J.~L. Robert, G.~E. Pikus, F.~G. Pikus, S.~V. Iordanskii V.~Mosser,
  K.~Zekentes, and Y.~B. Lyanda-Geller, {\it Weak antilocalization and spin
  precession in quantum wells\/}, Phys. Rev. B {\bf 53}, 3912 (1996).

\bibitem{lin2022}
J.-X. Lin, P.~Siriviboon, H.~D. Scammell, S.~Liu, D.~Rhodes, K.~Watanabe,
  T.~Taniguchi J.~Hone, M.~S. Scheurer, and J.~Li, {\it Zero-field
  superconducting diode effect in small-twist-angle trilayer graphene\/}, Nat.
  Phys. {\bf 18}, 1221 (2022).

\bibitem{brandt2004}
E.~H. Brandt and J.~R. Clem, {\it Superconducting thin rings with finite
  penetration depth\/}, Phys. Rev. B {\bf 69}, 184509 (2004).

\bibitem{mayer2020}
W.~Mayer, M.~C. Dartiailh, J.~Yuan K.~S. Wickramasinghe, E.~Rossi, and
  J.~Shabani, {\it Gate controlled anomalous phase shift in Al/{InAs} Josephson
  junctions\/}, Nat. Commun. {\bf 11}, 212 (2020).

\bibitem{clarke2004}
J.~Clarke and A.~I. Braginski, {\it The SQUID Handbook: Fundamentals and
  Technology of SQUIDs and SQUID Systems\/}, vol.~1 (Wiley-VCH, 2004).

\bibitem{Gennes1989}
P.~G. De~Gennes, {\it Superconductivity of {M}etals and {A}lloys\/}
  (Addison--Wesley, 1989).

\bibitem{Scharf2019}
B.~Scharf, F.~Pientka H.~Ren, A.~Yacoby, and E.~M. Hankiewicz, {\it Tuning
  topological superconductivity in phase-controlled {J}osephson junctions with
  {R}ashba and {D}resselhaus spin-orbit coupling\/}, Phys. Rev. B {\bf 99},
  214503 (2019).

\bibitem{lee2019}
J.~S. Lee, B.~Shojaei, M.~Pendharkar, A.~P. {McFadden}, Y.~Kim, H.~J. Suominen,
  M.~Kjaergaard, F.~Nichele H.~Zhang, C.~M. Marcus, and C.~J. Palmstrøm, {\it
  Transport Studies of Epi-{Al/InAs} Two-Dimensional Electron Gas Systems for
  Required Building-Blocks in Topological Superconductor Networks\/}, Nano
  Lett. {\bf 19}, 3083 (2019).

\bibitem{Clem2010}
J.~R. Clem, {\it Josephson junctions in thin and narrow rectangular
  superconducting strips\/}, Phys. Rev. B {\bf 81}, 144515 (2010).

\bibitem{Casimir1945}
H.~B.~G. Casimir, {\it On {O}nsager's Principle of Microscopic
  Reversibility\/}, Rev. Mod. Phys. {\bf 17}, 343 (1945).

\bibitem{Baumgartner_2022}
C.~Baumgartner, L.~Fuchs, A.~Costa, J.~Picó-Cortés, S.~Reinhardt, S.~Gronin,
  G.~C. Gardner, T.~Lindemann, M.~J. Manfra, P.~E.~F. Junior, D.~Kochan
  J.~Fabian, N.~Paradiso, and C.~Strunk, {\it Effect of {R}ashba and
  {D}resselhaus spin–orbit coupling on supercurrent rectification and
  magnetochiral anisotropy of ballistic {J}osephson junctions\/}, J. Phys.:
  Condens. Matter {\bf 34}, 154005 (2022).

\bibitem{daido2022}
A.~Daido, Y.~Ikeda, and Y.~Yanase, {\it Intrinsic Superconducting Diode
  Effect\/}, Phys. Rev. Lett. {\bf 128}, 037001 (2022).

\bibitem{Legg2022}
H.~F. Legg, D.~Loss, and J.~Klinovaja, {\it Superconducting diode effect due to
  magnetochiral anisotropy in topological insulators and {R}ashba nanowires\/},
  Phys. Rev. B {\bf 106}, 104501 (2022).

\bibitem{Dartiailh2021}
M.~C. Dartiailh, W.~Mayer, J.~Yuan, K.~S. Wickramasinghe A.~Matos-Abiague,
  I.~\ifmmode \check{Z}\else \v{Z}\fi{}uti\ifmmode~\acute{c}\else \'{c}\fi{},
  and J.~Shabani, {\it Phase signature of topological transition in {J}osephson
  junctions\/}, Phys. Rev. Lett. {\bf 126}, 036802 (2021).

\bibitem{Casparis2018}
L.~Casparis, M.~R. Connolly, M.~Kjaergaard, N.~J. Pearson, A.~Kringhøj, T.~W.
  Larsen, F.~Kuemmeth, T.~Wang, C.~Thomas, S.~Gronin, G.~C. Gardner M.~J.
  Manfra, C.~M. Marcus, and K.~D. Petersson, {\it Superconducting gatemon qubit
  based on a proximitized two-dimensional electron gas\/}, Nat. Nanotechnol.
  {\bf 13}, 915 (2018).

\bibitem{monroe2022}
D.~Monroe, M.~Alidoust, and I.~Žutić, {\it Tunable Planar Josephson Junctions
  Driven by Time-Dependent Spin-Orbit Coupling\/}, Phys. Rev. Applied {\bf 18},
  L031001 (2022).

\bibitem{monroe2024}
D.~Monroe, C.~Shen, D.~Tringali M.~Alidoust, T.~Zhou, and I.~Žutić, {\it
  Phase jumps in Josephson junctions with time-dependent spin–orbit
  coupling\/}, Appl. Phys. Lett. {\bf 125}, 012601 (2024).

\end{thebibliography}


\begin{thebibliography}{10}

\bibitem{shabani2016}
J.~Shabani, M.~Kjaergaard, H.~J. Suominen, Y.~Kim, F.~Nichele, K.~Pakrouski,
  T.~Stankevic, R.~M. Lutchyn, P.~Krogstrup, R.~Feidenhans'l, S.~Kraemer,
  C.~Nayak M.~Troyer, C.~M. Marcus, and C.~J. Palmstrøm, {\it Two-dimensional
  epitaxial superconductor-semiconductor heterostructures: A platform for
  topological superconducting networks\/}, Phys. Rev. B {\bf 93}, 155402
  (2016).

\bibitem{kjaergaard2016}
M.~Kjaergaard, F.~Nichele, H.~J. Suominen, M.~P. Nowak, M.~Wimmer, A.~R.
  Akhmerov, J.~A. Folk, K.~Flensberg J.~Shabani, C.~J. Palmstrøm, and C.~M.
  Marcus, {\it Quantized conductance doubling and hard gap in a two-dimensional
  semiconductor–superconductor heterostructure\/}, Nat. Commun. {\bf 7},
  12841 (2016).

\bibitem{knap1996}
W.~Knap, C.~Skierbiszewski, A.~Zduniak, E.~Litwin-Staszewska, D.~Bertho,
  F.~Kobbi, J.~L. Robert, G.~E. Pikus, F.~G. Pikus, S.~V. Iordanskii V.~Mosser,
  K.~Zekentes, and Y.~B. Lyanda-Geller, {\it Weak antilocalization and spin
  precession in quantum wells\/}, Phys. Rev. B {\bf 53}, 3912 (1996).

\bibitem{lin2022}
J.-X. Lin, P.~Siriviboon, H.~D. Scammell, S.~Liu, D.~Rhodes, K.~Watanabe,
  T.~Taniguchi J.~Hone, M.~S. Scheurer, and J.~Li, {\it Zero-field
  superconducting diode effect in small-twist-angle trilayer graphene\/}, Nat.
  Phys. {\bf 18}, 1221 (2022).

\bibitem{brandt2004}
E.~H. Brandt and J.~R. Clem, {\it Superconducting thin rings with finite
  penetration depth\/}, Phys. Rev. B {\bf 69}, 184509 (2004).

\bibitem{Dartiailh2021}
M.~C. Dartiailh, W.~Mayer, J.~Yuan, K.~S. Wickramasinghe A.~Matos-Abiague,
  I.~\ifmmode \check{Z}\else \v{Z}\fi{}uti\ifmmode~\acute{c}\else \'{c}\fi{},
  and J.~Shabani, {\it Phase signature of topological transition in {J}osephson
  junctions\/}, Phys. Rev. Lett. {\bf 126}, 036802 (2021).

\bibitem{mayer2020}
W.~Mayer, M.~C. Dartiailh, J.~Yuan K.~S. Wickramasinghe, E.~Rossi, and
  J.~Shabani, {\it Gate controlled anomalous phase shift in Al/{InAs} Josephson
  junctions\/}, Nat. Commun. {\bf 11}, 212 (2020).

\bibitem{clarke2004}
J.~Clarke and A.~I. Braginski, {\it The SQUID Handbook: Fundamentals and
  Technology of SQUIDs and SQUID Systems\/}, vol.~1 (Wiley-VCH, 2004).

\bibitem{Souto2022}
R.~S. Souto, M.~Leijnse, and C.~Schrade, {\it Josephson Diode Effect in
  Supercurrent Interferometers\/}, Phys. Rev. Lett. {\bf 129}, 267702 (2022).

\bibitem{ciaccia2023}
C.~Ciaccia, R.~Haller, A.~C.~C. Drachmann, T.~Lindemann M.~J. Manfra,
  C.~Schrade, and C.~Schönenberger, {\it Gate-tunable Josephson diode in
  proximitized {InAs} supercurrent interferometers\/}, Phys. Rev. Research {\bf
  5}, 033131 (2023).

\bibitem{Gennes1989}
P.~G. De~Gennes, {\it Superconductivity of {M}etals and {A}lloys\/}
  (Addison--Wesley, 1989).

\bibitem{Baumgartner_2022}
C.~Baumgartner, L.~Fuchs, A.~Costa, J.~Picó-Cortés, S.~Reinhardt, S.~Gronin,
  G.~C. Gardner, T.~Lindemann, M.~J. Manfra, P.~E.~F. Junior, D.~Kochan
  J.~Fabian, N.~Paradiso, and C.~Strunk, {\it Effect of {R}ashba and
  {D}resselhaus spin–orbit coupling on supercurrent rectification and
  magnetochiral anisotropy of ballistic {J}osephson junctions\/}, J. Phys.:
  Condens. Matter {\bf 34}, 154005 (2022).

\bibitem{Scharf2019}
B.~Scharf, F.~Pientka H.~Ren, A.~Yacoby, and E.~M. Hankiewicz, {\it Tuning
  topological superconductivity in phase-controlled {J}osephson junctions with
  {R}ashba and {D}resselhaus spin-orbit coupling\/}, Phys. Rev. B {\bf 99},
  214503 (2019).

\bibitem{lee2019}
J.~S. Lee, B.~Shojaei, M.~Pendharkar, A.~P. {McFadden}, Y.~Kim, H.~J. Suominen,
  M.~Kjaergaard, F.~Nichele H.~Zhang, C.~M. Marcus, and C.~J. Palmstrøm, {\it
  Transport Studies of Epi-{Al/InAs} Two-Dimensional Electron Gas Systems for
  Required Building-Blocks in Topological Superconductor Networks\/}, Nano
  Lett. {\bf 19}, 3083 (2019).

\bibitem{davydova2022}
M.~Davydova, S.~Prembabu, and L.~Fu, {\it Universal {J}osephson diode
  effect\/}, Sci. Adv. {\bf 8}, eabo0309 (2022).

\bibitem{Clem2010}
J.~R. Clem, {\it Josephson junctions in thin and narrow rectangular
  superconducting strips\/}, Phys. Rev. B {\bf 81}, 144515 (2010).

\bibitem{banerjee2023}
A.~Banerjee, M.~Geier, M.~A. Rahman, C.~Thomas, T.~Wang M.~J. Manfra,
  K.~Flensberg, and C.~M. Marcus, {\it Phase Asymmetry of {A}ndreev Spectra
  from {C}ooper-Pair Momentum\/}, Phys. Rev. Lett. {\bf 131}, 196301 (2023).

\bibitem{yuan2022}
N.~F.~Q. Yuan and L.~Fu, {\it Supercurrent diode effect and finite-momentum
  superconductors\/}, Proc. Natl. Acad. Sci. USA {\bf 119}, e2119548119 (2022).

\bibitem{Beenakker1991}
C.~W.~J. Beenakker, {\it Universal limit of critical-current fluctuations in
  mesoscopic {J}osephson junctions\/}, Phys. Rev. Lett. {\bf 67}, 3836 (1991).

\end{thebibliography}
\bibliographystyle{JDE}

\end{document}


\baselineskip18pt
\maketitle 

\section{Characteristics of Al-InAs heterostructure}

\begin{figure}[h!]%
\centering
\includegraphics[width=0.8\textwidth]{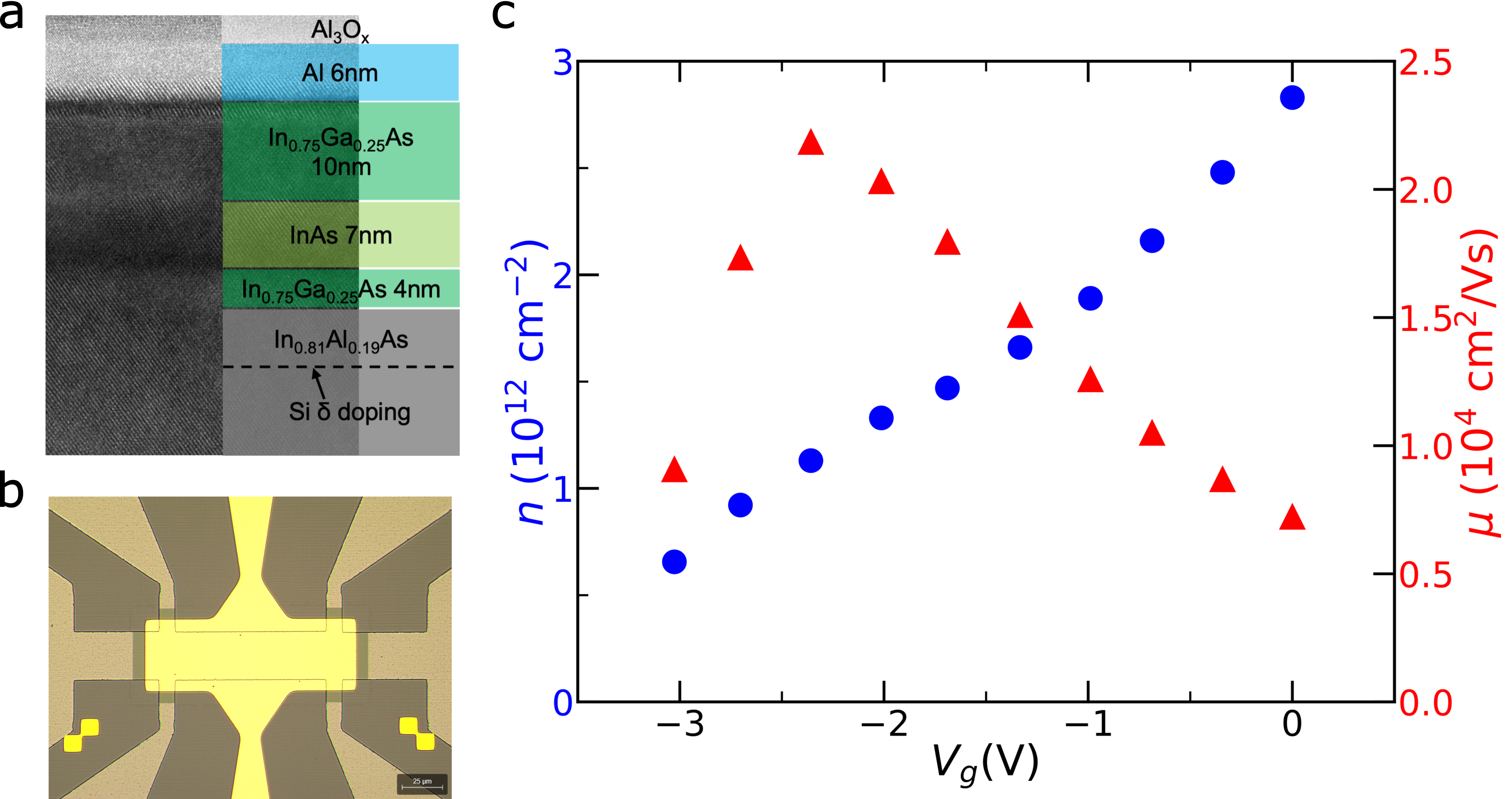}
\caption{Structure and transport characteristics of an InAs quantum well. 
(a) Cross-sectional transmission electron microscopy (TEM) image showing the layered configuration of an Al-InAs heterostructure. 
(b) Optical image of a Hall-bar-shaped device with the InAs quantum well with a top gate. 
(c) Electron density ($n$) and mobility ($\mu$) as a function of gate voltage ($V_{g}$) in the InAs quantum well.}
\label{figS1}
\end{figure}

The Al-InAs heterostructure is grown on a semi-insulating InP substrate by molecular beam epitaxy (Fig.~\ref{figS1}(a)). The heterostructure consists of a 100 nm $\text{In}_{0.52} \text{Al}_{0.48} \text{As}$ matched buffer, 25 nm $\text{In}_{0.52} \text{Al}_{0.48} \text{As}/\text{In}_{0.52} \text{Ga}_{0.48} \text{As}$ superlattice, 800 nm $\text{In}_{x} \text{Al}_{1-x} \text{As}$ graded buffer ($x$ = 0.52 to 0.81), 25 nm $\text{In}_{0.81} \text{Al}_{0.19} \text{As} /\text{In}_{0.75} \text{Ga}_{0.25} \text{As}$ superlattice, 106 nm $\text{In}_{0.81} \text{Al}_{0.19} \text{As}$ the topmost buffer layer, a 4-nm-thick $\text{In}_{0.75} \text{Ga}_{0.25} \text{As}$ bottom barrier, a 7 nm InAs quantum well, a 10 nm $\text{In}_{0.75} \text{Ga}_{0.25} \text{As}$ top barrier, and 6 nm aluminum film as a superconducting layer. A Si-$\delta$-doping layer with a sheet density of $1.4 \times 10^{12}$ $\text{cm}^{-2}$ is placed 6 nm below the top surface of the buffer layer. The Al layer is epitaxially grown without breaking vacuum to provide a clean proximity contact to the underlying quantum well, inducing a superconducting gap in the InAs layer comparable to that of the Al layer \cite{shabani2016,kjaergaard2016}.

\begin{figure}[h!]%
\centering
\includegraphics[width=1\textwidth]{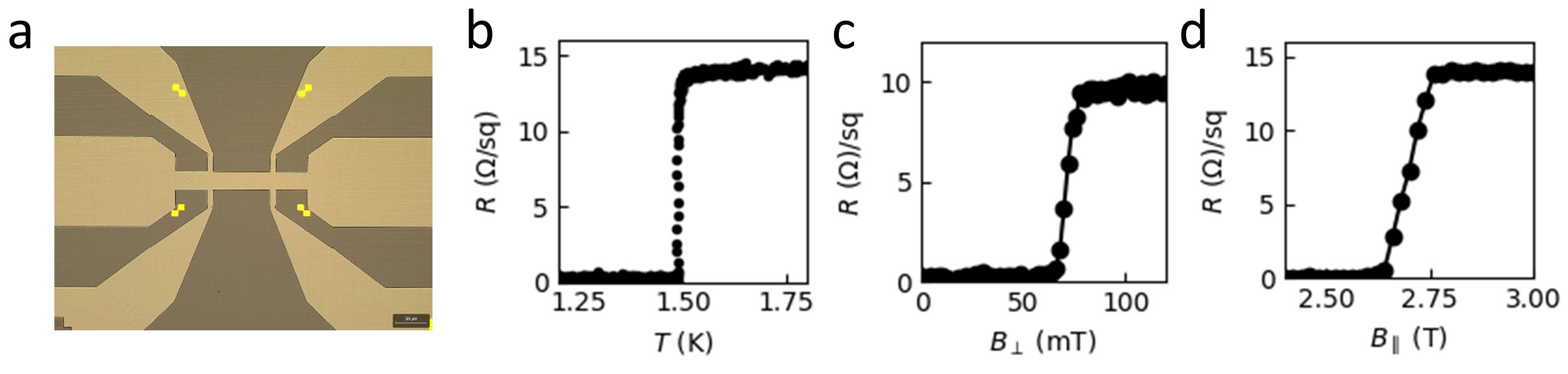}
\caption{
Characteristics of the top superconducting Al layer. 
(a) Optical image of a Hall-bar-shaped Al film grown on the InAs quantum well.
(b) Resistance ($R$) as a function of temperature ($T$) at zero external magnetic field. 
(c) $R$ as a function of the out-of-plane magnetic field ($B_{\bot}$).
(d) $R$ as a function of the in-plane magnetic field ($B_{\parallel}$).
}
\label{figS3}
\end{figure}

We perform a characterization of the InAs quantum well in a top-gated Hall-bar geometry device where the top Al layer is removed (Fig.~\ref{figS1}(b)). The characterization of the quantum well reveals a peak mobility of 2.2$\times 10^{4}$ $\text{cm}^{2}/\text{Vs}$ at an electron sheet density of 1.1$\times 10^{12}$ $\text{cm}^{-2}$ (Fig.~\ref{figS1}(c)), corresponding to an electron mean free path $l_{e} \approx$ 380 nm. Meanwhile, we perform a characterization of the Al layer in a Hall-bar geometry device (Fig.~\ref{figS3}). The zero-field transition temperature is 1.49 K and the in-plane critical magnetic field is 2.7 T at 10 mK. Using the relation $\Delta = 1.75 k_{B} T_{c}$, $\Delta$ is estimated to be approximately 200 $\upmu\text{eV}$. 

\newpage

\section{Spin-orbit couplings in InAs quantum well}

An InAs quantum well exhibits two types of SOCs: Rashba and Dresselhaus. Dresselhaus SOC arises due to the lack of inversion symmetry in the crystal structure. In bulk semiconductors with a zinc blende structure, such as GaAs, InAs, and InSb, this broken inversion symmetry naturally leads to cubic Dresselhaus SOC, denoted as $\gamma$. In a quantum well, the confinement modifies the electron wave function so that the electron's motion is primarily in the plane of the quantum well. The quantization along the growth direction (z-direction) alters the form of the Dresselhaus SOC, resulting in a linear Dresselhaus SOC, $\beta$. Rashba SOC, represented by $\alpha$, is induced by structural inversion asymmetry, and its strength depends on the net electric field. In a quantum well, an electric field in the z-direction can arise due to the combined effects of structural inversion asymmetry and the mixing of valence band states with conduction band states. This electric field can be further controlled by applying an external electric field through gating.

\begin{figure}[h!]%
\centering
\includegraphics[width=0.8\textwidth]{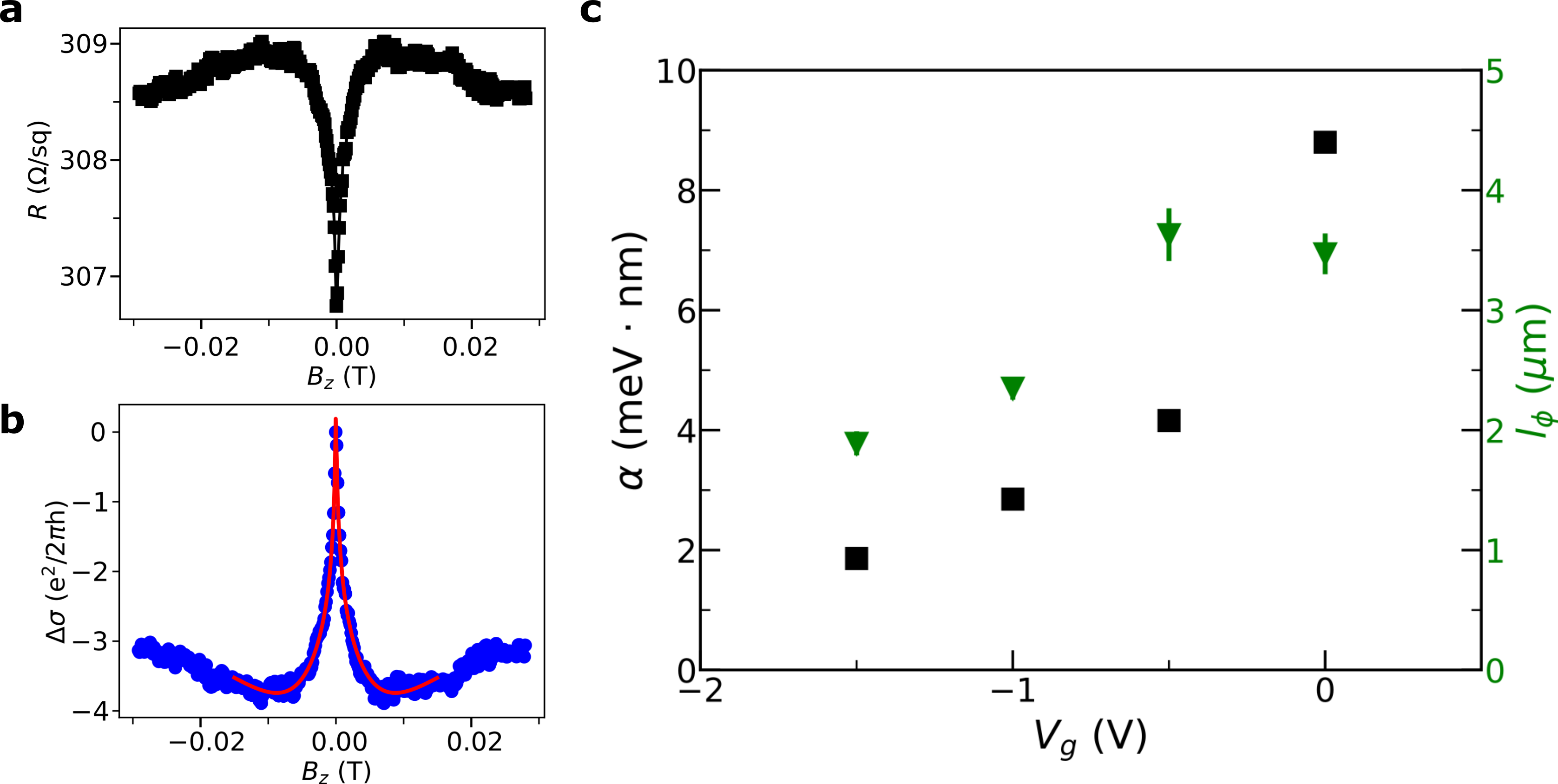}
\caption{
Magnetoconductance analysis and Rashba SOC.
(a) Resistance ($R$) as a function of the perpendicular magnetic fields ($B_z$) at $V_{g}$ = 0 V.
(b) Magnetoconductance variation ($\Delta\sigma = \sigma(B_{z})-\sigma(0)$) plotted as a function of $B_z$, with blue circles representing the experimental data and the red line indicating the fit using the ILP model.
(c) Rashba SOC $\alpha$ and phase-coherence length $l_{\phi}$ as a function of $V_{g}$.
}
\label{figS2}
\end{figure}

To examine the evolution of Rashba SOC in the InAs quantum well as a function of gate voltage, we measure a weak antilocalization signal using the Hall-bar device and extract Rashba SOC, $\alpha$, by analyzing the signal (Fig.~\ref{figS2}). We use the theory developed by Iordanski, Lyanda-Geller, and Pikus (ILP) for a 2D electron gas, which is valid when either Rashba SOC or linear Dresselhaus SOC is dominant \cite{shabani2016, knap1996}. 
\begin{equation}
\begin{split}
    \Delta\sigma(B) = & -\frac{e^2}{4\pi^2\hbar}\biggl\{\frac{1}{a_0}+\frac{2a_0+1+(H_{SO}/B)}{a_1[a_0+(H_{SO}/B)]-(2H'_{SO}/B)} \\
     & -\sum_{n=1}^{\infty}\left[\frac{3}{n}-\frac{3a_{n}^{2}+2a_{n}(H_{SO}/B)-1-2(2n+1)(H'_{SO}/B)}{a_{n}+(H_{SO}/B)a_{n-1}a_{n+1}-2(H'_{SO}/B)[(2n+1)a_{n}-1]}\right] \\
     & + 2\ln{\frac{H_{tr}}{B}} + \Psi(\frac{1}{2}+\frac{H_{\varphi}}{B})+3C\biggl\}\\
    &   \\
    & a_{n} = n + \frac{1}{2} + \frac{H_{\varphi}}{B} + \frac{H_{SO}}{B}\text{, } H_{\varphi} = \frac{\hbar}{4el_{\phi}^{2}}\text{, } H_{tr} = \frac{\hbar}{4eD\tau_{e}}\text{, }\\ 
    & H_{SO} = H_{SO1} + H_{SO3}\text{, }
    H_{SO1} = H_{\alpha} + H_{\beta}\text{, } H'_{SO} = H_{\alpha} \text{ or } H_{\beta}\text{, }\\
    & H_{SO3} = \frac{1}{4eD\hbar}(2\Omega_{\gamma}^2\tau_e)\text{, } H_{\alpha} = \frac{1}{4eD\hbar}(2\Omega_{\alpha}^2\tau_e)\text{, }  H_{\beta} = \frac{1}{4eD\hbar}(2\Omega_{\beta}^2\tau_e)\text{, } \\
    &\Omega_{\gamma} = \gamma(k_{f}^3)/4\text{, }
    \Omega_{\alpha} = \alpha k_{f}\text{, }
    \Omega_{\beta} = \beta k_{f}\text{, }
\end{split}
\end{equation}
where, $\Psi$ is digamma-function, $C$ is Euler's constant, $k_{f}$ is the Fermi wave vector, and $\tau_e$ is the elastic scattering time. In the analysis, we use an effective electron mass of $m^{*} = 0.023m_{e}$ and a cubic Dresselhaus SOC value of $\gamma = 2.69 \times 10^{-2}$ $\text{eV} \cdot \text{nm}^3$, calculated from $\overrightarrow{k} \cdot \overrightarrow{p}$ theory \cite{knap1996}. Additionally, we use $\beta = 4.23$  $\text{meV} \cdot \text{nm}$, a value employed for fitting the $B_y$ dependence of the diode efficiency (see Table~\ref{tab:parameters}). The estimated $\beta$ for similar InAs quantum wells in Ref.~\cite{shabani2016} is 5 $\text{meV} \cdot \text{nm}$, which is comparable to the value of $\beta$ we used.

From fitting the weak antilocalization signal, we extract the Rashba SOC, $\alpha$, and the phase-coherence length, $l_{\phi}$. At $V_{g} = 0$, the obtained $\alpha$ from the weak antilocalization signal is 8.8 $\text{meV} \cdot \text{nm}$, which is comparable to $\alpha = 7.53$  $\text{meV} \cdot \text{nm}$ obtained from fitting the diode efficiency (see Figure 4a in the main text and Table~\ref{tab:parameters}). 

The $\alpha$ in the Hall-bar sample is suppressed when a negative gate voltage is applied (Fig.~\ref{figS2}(c)). This is more pronounced than the one obtained from fitting the diode efficiency (see Sec.\ref{fit parameters}). We attribute the discrepancy to the difference in effective gate voltage between the Hall-bar-shaped device with a wide gating region and the superconducting quantum interference device (SQUID) with narrow gating regions. 

\section{Fabrication}
The fabrication process for the devices involves several standard electron-beam lithography steps. These include defining MESAs and Josephson junctions and depositing top gate electrodes. MESAs are isolated by a 270 nm chemical wet etch using a solution ($\text{H}_{2}\text{O}:\text{C}_{6}\text{H}_{8}\text{O}_{7}:\text{H}_{3}\text{PO}_{4}:\text{H}_{2}\text{O}_{2}$ = 220:55:3:3) after the local removal of Al layers using a wet etchant (Transene D). Josephson junctions on the MESAs are defined by selectively removing Al layers. Subsequently, a 20-nm-thick insulating Al$_{\rm 2}$O$_{\rm x}$ layer is deposited via atomic layer deposition. The top gate electrodes are deposited in two steps using electron-beam evaporation: 5 nm Ti and 30 nm Au deposition for fine structures, followed by a successive deposition of 10 nm Ti and 300 nm Au deposition for larger structures.

\section{Measurements}
All electrical measurements are conducted in a dilution refrigerator equipped with suitable electronic low-pass filters at the mixing chamber stage, with a base temperature of 10 mK. Two different direction magnetic fields, in-plane field $B_{y}$ and out-of-plane field $B_{z}$, are applied by a superconducting magnet installed in the refrigerator and by a homemade superconducting coil attached to the lid of a sample holder, respectively. Due to misalignment between the applied $B_y$ field and the sample surface, an unintended $B_z$ component is introduced. To compensate for this, an additional $B_z$ offset field is applied using the homemade superconducting coil. By appropriately tuning this offset, the experiment is performed near the maximum of the Fraunhofer-like modulated SQUID oscillation (see Sec.~\ref{characSQUID}), ensuring that the actual $B_z$ field remains close to zero.

DC current-biased differential resistance (d$V$/d$I$) measurements are performed by using standard low-frequency lock-in techniques with an excitation current $I_{ex}$ = 10 nA. To obtain forward (backward) critical currents $I_c^{+}$ ($I_c^-$), DC bias current $I_{DC}$ sweeps increasing (decreasing) from zero. In cases with an abrupt d$V$/d$I$ jump, as shown in Figure 2b in the main text, the current value at the abrupt resistance jump is extracted as the forward (backward) critical current. With increasing $\lvert B_{y} \rvert$ or applying negative gate voltages, the discontinuous d$V$/d$I$($I_{DC}$) changes to a continuous curve with a d$V$/d$I$ peak. We define the critical currents in the continuous curves where d$V$/d$I$ reaches its maximum value. This definition is consistent with that used in Ref.~\cite{lin2022}

\newpage
\section{Characteristics of Al-InAs SQUID}\label{characSQUID}
The SQUID consists of two Josephson junctions (JJs), each with a nominal junction length of 100 nm ($L_j$) and a width of 4.5 $\upmu$m ($W$). The superconducting leads connected to each junction include a short segment of approximately 1.4 $\upmu$m and a longer extension of about 4.4 $\upmu$m that incorporates the SQUID arm. The enclosed area of the SQUID loop ($A_{\text{SQUID}}$) is 20~$(\upmu\text{m})^2$.

\begin{figure}[h!]%
\centering
\includegraphics[width=0.8\textwidth]{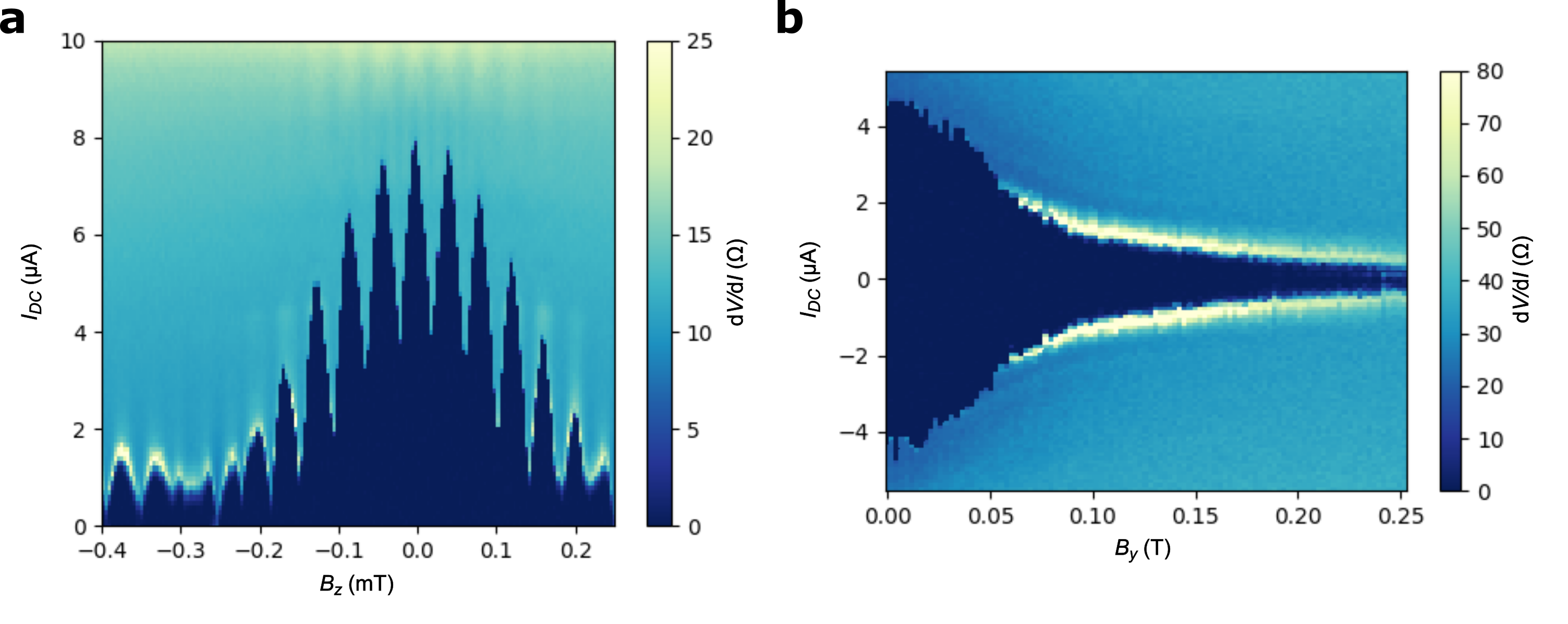}
\caption{
Al-InAs SQUID. 
(a) Differential resistance (d$V$/d$I$) as a function of $B_z$ and $I_{DC}$ at $B_y = 0$ and $V_{g1} = V_{g2} = 0$. 
(b) Differential resistance as a function of $B_y$ and $I_{DC}$ at $B_z = 0$, with $V_{g1} = -15$ V and $V_{g2} = 0$.
}
\label{figS4}
\end{figure}

Figure~\ref{figS4}(a) shows the differential resistance (d$V$/d$I$) as a function of $B_z$ (ranging from $-0.4$ to 0.25 mT) and $I_{DC}$, measured at $B_y = 0$ and $V_{g1} = V_{g2} = 0$. This figure reveals SQUID oscillations with a period of $4.2\times 10^{-5}$ T, modulated by a Fraunhofer-like pattern in the Josephson junctions (JJs). The oscillation period is approximately half of $\Phi_0 / A_{\text{SQUID}}$, rather than $\Phi_0 / A_{\text{SQUID}}$. Here, $A_{\text{SQUID}}$ is the enclosed area of the SQUID loop. This is attributed to the larger effective area resulting from magnetic field focusing due to the Meissner effect in the wide SQUID ring \cite{brandt2004}.

Figure~\ref{figS4}(b) displays the differential resistance (d$V$/d$I$) as a function of $B_y$ and $I_{DC}$, measured at $B_z = 0$, with $V_{g1} = -15$ V and $V_{g2} = 0$. This gate voltage configuration shows the dependence of the critical current in J2, while J1 is nearly pinched off. The critical current of J2 gradually diminishes with increasing $B_y$, reaching approximately half its original value near $B_y = 50$ mT. Across the range of 0 to 0.25 T, no nonmonotonic behavior (such as suppression and then revival of the critical current) is observed, which is interpreted as a potential indicator of a topological phase transition \cite{Dartiailh2021}.\\

\newpage
\section{Nonreciprocal critical currents in an Al-InAs SQUID}
In SQUIDs, nonreciprocal critical current can arise from having non-identical junctions within the device, even if each junction does not exhibit nonreciprocal properties. 
In contrast, this study investigates a different source of nonreciprocal critical current: the Josephson diode effect (JDE), where the JJs themselves exhibit nonreciprocal behavior. 

We numerically simulate the oscillations of the critical current in a SQUID under two distinct scenarios: one where nonreciprocity arises from junction non-identicality and another where it originates from the JDE. These simulations illustrate how the two sources of nonreciprocal critical currents influence a SQUID oscillation differently.

\bigskip
\subsection{Critical current oscillations in a SQUID}

\begin{figure}[h]%
\centering
\includegraphics[width=0.9\textwidth]{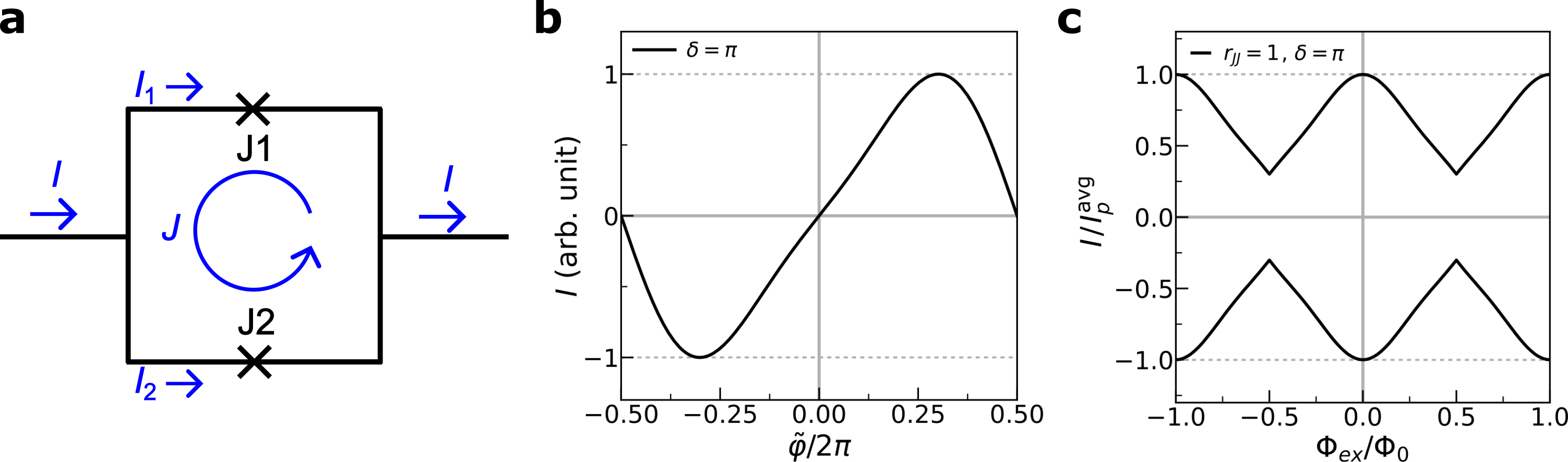}
\caption{
SQUID oscillation.
(a) Schematic diagram of a SQUID.
(b) CPR in a JJ with $a_{1} = 1$, $a_{2}/a_{1} = 0.2$, and $\delta = \pi$, which corresponds to the black solid line in Fig.~1(c) of the main text.
(c) Maximum and minimum supercurrent in a SQUID, with $a_{1} = 1$, $a_{2}/a_{1} = 0.2$, $\delta = \pi$, and $r_{JJ} = 1$, which corresponds to the black solid line in Fig.~1(d) of the main text. $I_{p}^{\text{avg}} = \left(I_{p}^{+}-I_{p}^{-}\right)/2$, where $I_{p}^{+}$ and $I_{p}^{-}$ are the maximum magnitudes of the maximum and minimum supercurrent oscillations, respectively.  
}
\label{figS5}
\end{figure}

\noindent To calculate the critical current in a SQUID, we consider a DC SQUID composed of two JJs, J1 and J2 (Fig.~\ref{figS5}(a)). We assume the current-phase relations (CPRs) of the JJs include second harmonics, expressed as $I \left(\varphi\right) = a_{1}\sin{\left(\varphi + \varphi_{1}\right)} + a_{2}\sin{\left(2\varphi + \varphi_{2}\right)}$, where $\varphi$ represents the Josephson phase, $a_1$ and $a_2$ are the amplitudes of the first and second harmonics, and $\varphi_{1}$ and $\varphi_{2}$ are the phase offsets of the first and second harmonics. This equation can be alternatively expressed as $I \left(\tilde{\varphi}\right) = a_{1}\sin{\left(\tilde{\varphi}\right)} + a_{2}\sin{\left(2\tilde{\varphi} + \delta\right)}$, where $\delta = \varphi_{2}-2\varphi_{1}$. This form is the same as Eq. (1) in the main text. The ratio $a_{2}/a_{1}$ and the phase difference $\delta$ determine the efficiency of the JDE, with $\delta$ influencing its polarity. For simplicity, we assume that both JJs have identical nonreciprocal characteristics, characterized by the same $\delta$ and $a_{2}/a_{1}$. Then, the CPRs for J1 and J2 are: 
\begin{equation}
    I_{1}(\tilde{\varphi}^{\text{J1}}) = a_{1}\sin{\left(\tilde{\varphi}^{\text{J1}}\right)}+a_{2}\sin{\left(2\tilde{\varphi}^{\text{J1}}+\delta\right)} \label{CPR-J1}
\end{equation}
\begin{equation}
    I_{2}(\tilde{\varphi}^{\text{J2}}) = r_{JJ}\left(a_{1}\sin{\left(\tilde{\varphi}^{\text{J2}}\right)}+a_{2}\sin{\left(2\tilde{\varphi}^{\text{J2}}+\delta\right)}\right)
\end{equation},
where the superscripts indicate the corresponding JJs. We introduce the parameter $r_{JJ}$ into the CPR for J2 to account for the difference in magnitudes between the two junctions.

Figure~\ref{figS5}(a) shows a schematic diagram of a SQUID, where $I$ is the total current passing through the SQUID, and $J$ is the circulating current in the SQUID loop. The currents passing through JJs $I_{1}$ and $I_{2}$ can be expressed as $I_{1} = I/2 - J$ and $I_{2} = I/2 + J$, respectively. Therefore, $I$ and $J$ follow these relationships: 
 \begin{equation}\label{eq.S4}
     I = I_{1} + I_{2}
 \end{equation}
and 
\begin{equation}\label{eq.S5}
    J = \frac{1}{2}\left(I_{2} - I_{1}\right).
\end{equation}
Due to fluxoid quantization, the Josephson phases satisfy the following relation:
\begin{equation}\label{eq.S6}
    \varphi^{\text{J2}}-\varphi^{\text{J1}}+\left(2\pi/\Phi_{0}\right)\left(\Phi_{ex}+LJ\right)=2\pi n
\end{equation}
, where $\Phi_{ex}$ is the external flux, $L$ is the SQUID inductance and $n$ is an integer.
Using Eq.~\ref{eq.S5}, this can be rewritten as
\begin{equation}\label{eq.S7}
    \varphi^{\text{J2}}-\varphi^{\text{J1}}+\left(2\pi/\Phi_{0}\right)\left(\Phi_{ex}+\frac{1}{2}L(I_{2}-I_{1})\right)=2\pi n.
\end{equation}
In the numerical calculation, we set the SQUID inductance to 0.18 nH, which provides the best fit to the experimental data (Sec.~\ref{fit}).

To estimate the critical current of a SQUID, we first find numerous combinations of $\varphi^{\text{J1}}$ and $\varphi^{\text{J2}}$ that satisfy Eq.~\ref{eq.S7} and compute the corresponding $I$ values using Eq.~\ref{eq.S4}. The maximum $I$ value among these is taken as the forward critical current ($I_c^{+}$), and the minimum value is considered the backward critical current ($I_c^{-}$).

Before discussing the results of the SQUID oscillation computations, we will examine the phase offset in the first harmonic, $\varphi_{1}$. When $\varphi_{1}^{J1} \neq 0$ ($\varphi_{1}^{J2} \neq 0$), the SQUID oscillation is shifted along the $\Phi_{ex}/\Phi_{0}$ axis by $\varphi_{1}^{J1}$ ($-\varphi_{1}^{J2}$) \cite{mayer2020}. Thus, the SQUID oscillation is shifted by $\varphi_{1}^{J1} - \varphi_{1}^{J2}$. We assume $\varphi_{1}^{J1} = \varphi_{1}^{J2}$, resulting in no shift in the SQUID oscillation in our computations.  

Figure~\ref{figS5}(c) shows an example of the calculation results for the case where no supercurrent rectification effect is present, with $\delta = \pi$ and $r_{JJ} = 1$. When $\delta = \pi$, there is no JDE, as shown in the CPR in Fig.~\ref{figS5}(b). The parameter $r_{JJ} = 1$ indicates that there is no discrepancy between the CPRs of J1 and J2. In this case, the oscillating critical currents are symmetric about both the $\Phi_{ex}$-axis and the $I$-axis, resulting in the forward and backward critical currents having the same value for all external flux.\\
\bigskip
\bigskip
\bigskip

\newpage

\bigskip
\subsection{Nonreciprocity due to non-identical JJs}

\begin{figure}[h]%
\centering
\includegraphics[width=0.8\textwidth]{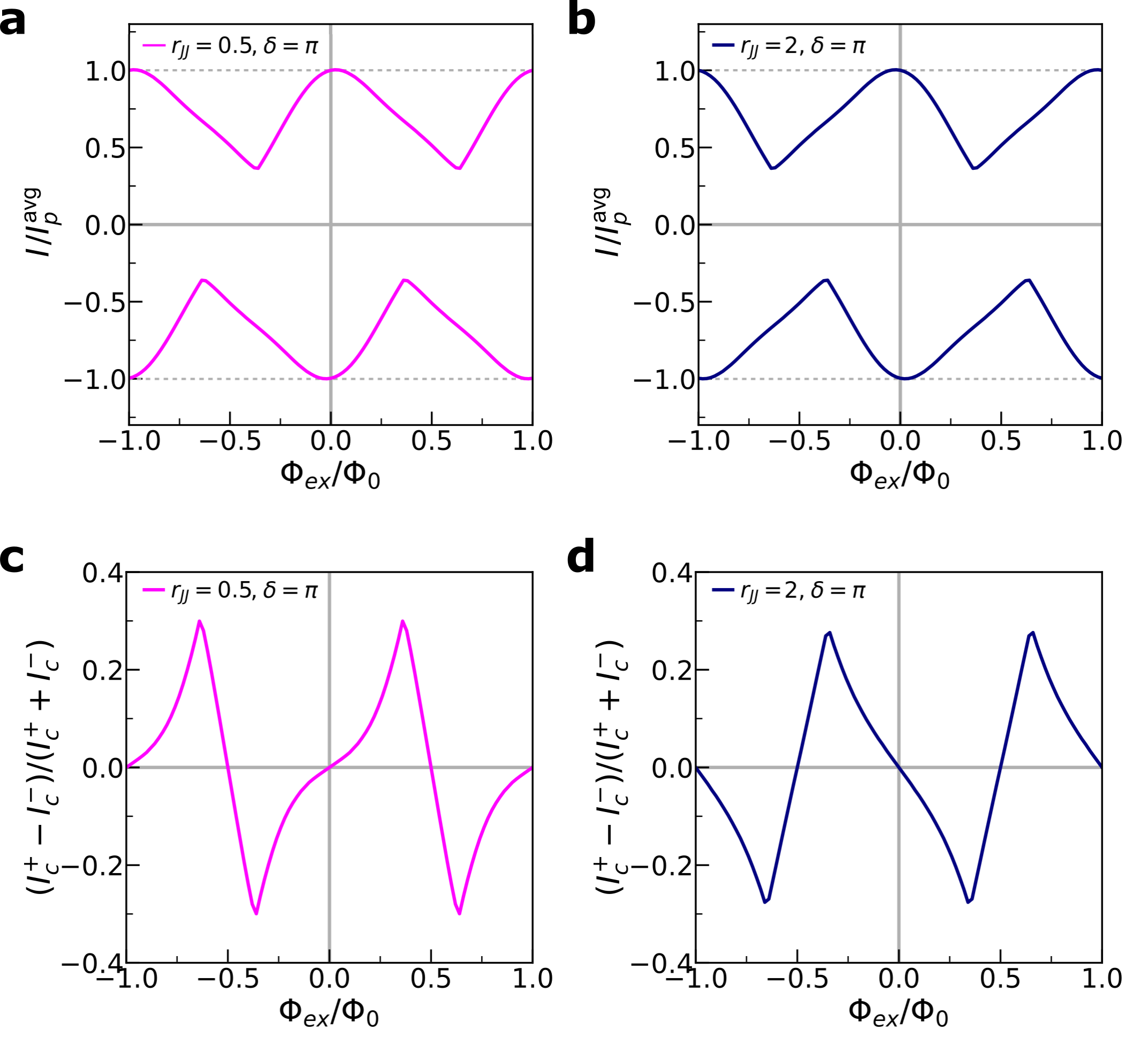}
\caption{
Critical current oscillation of a SQUID with non-identical JJs. 
(a) Maximum and minimum supercurrent in a SQUID, with $a_{1} = 1$, $a_{2}/a_{1} = 0.2$, $\delta = \pi$, and $r_{JJ} = 0.5$. 
(b) Maximum and minimum supercurrent in a SQUID, with $a_{1} = 1$, $a_{2}/a_{1} = 0.2$, $\delta = \pi$, and $r_{JJ} = 2$.
(c) $(I_{c}^{+} - I_{c}^{-})/(I_{c}^{+} + I_{c}^{-})$ as a function of $\Phi_{ex}/\Phi_{0}$ for the case with $a_{1} = 1$, $a_{2}/a_{1} = 0.2$, $\delta = \pi$ and $r_{JJ} = 0.5$. 
(d) $(I_{c}^{+} - I_{c}^{-})/(I_{c}^{+} + I_{c}^{-})$ as a function of $\Phi_{ex}/\Phi_{0}$, for the case with $a_{1} = 1$, $a_{2}/a_{1} = 0.2$, $\delta = \pi$, and $r_{JJ} = 2$.  
}
\label{figS6}
\end{figure}

\noindent We demonstrate the nonreciprocal supercurrent effect due to non-identical JJs in a SQUID, achieved by setting $r_{JJ} \neq 1$. When $r_{JJ} \neq 1$, the critical currents of the two JJs differ, despite all other characteristics being identical. Figure~\ref{figS6} shows examples with $r_{JJ} = 0.5$ and $r_{JJ} = 2$, with the remaining parameters, including $\delta$, are the same as in the SQUID oscillation case shown in Fig.~\ref{figS5}(c). The SQUID oscillation is no longer symmetric about the $\Phi_{ex}$-axis and the $I$-axis. This deformation of the SQUID oscillation leads to a nonreciprocal critical current. In Figs.~\ref{figS6}(c) and \ref{figS6}(d), the nonreciprocal diode effect reaches its maximum at $\Phi_{ex}/\Phi_{0} = \pm 0.34$ and $\pm 0.66$. To observe this nonreciprocal supercurrent effect, the device must meet specific conditions, such as having a non-zero SQUID inductance or higher harmonics in the CPR of the JJs\cite{clarke2004, Souto2022, ciaccia2023}.

Despite the asymmetry about the $\Phi_{ex}$-axis and $I$-axis, the anti-symmetry condition $\lvert I_c^+ \rvert (\Phi_{ex}) = -
\lvert I_c^-\rvert (-\Phi_{ex})$ remains. This ensures that the maximum magnitudes of the forward and backward critical currents, $I_p^+$ and $I_p^-$, are still equal. This behavior differs from the nonreciprocal SQUID oscillations driven by the JDE (Sec.~\ref{subsec3}).

\begin{figure}[h]%
\centering
\includegraphics[width=0.8\textwidth]{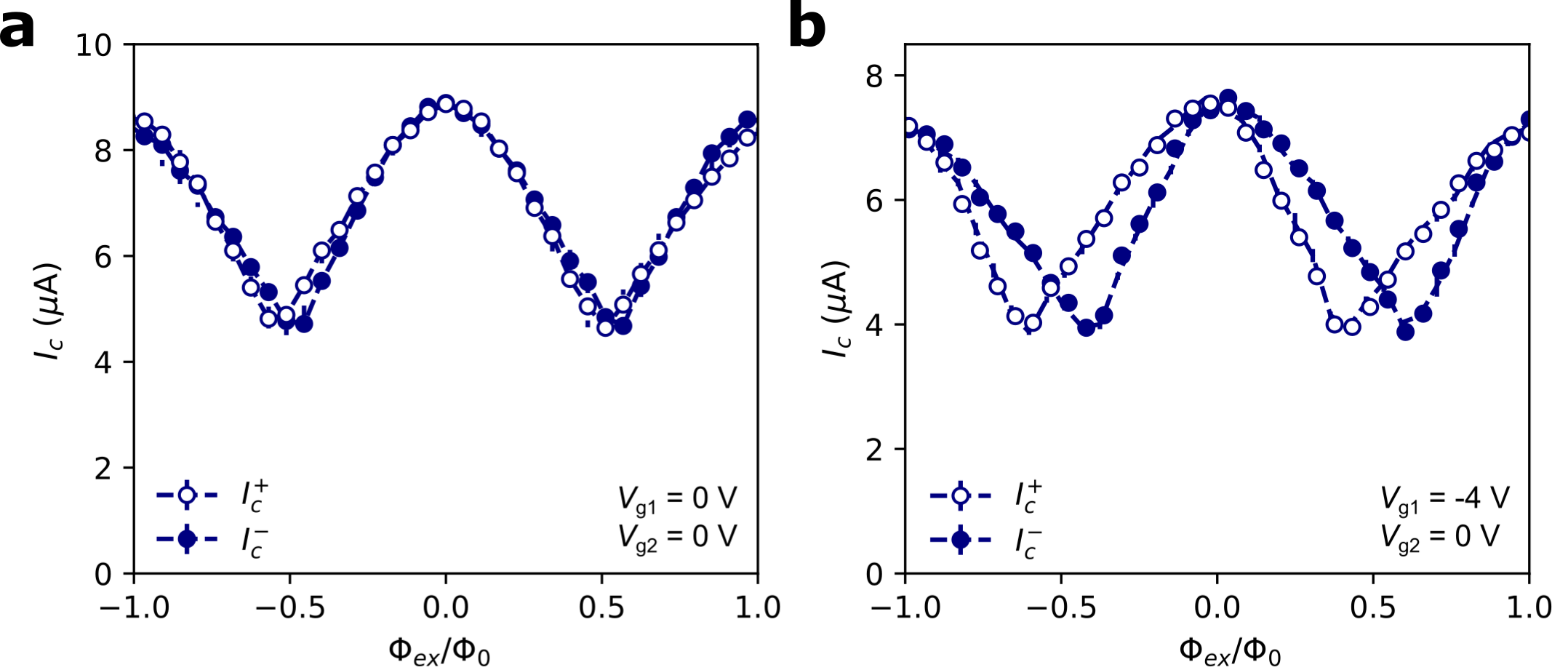}
\caption{
Critical current oscillation of the SQUID with different gate voltage combinations.
(a) SQUID oscillation of forward and backward critical currents with $V_{g1} = 0$ V and $V_{g2} = 0$ V.
(b) QUID oscillation of forward and backward critical currents with $V_{g1} = -4$ V and $V_{g2} = 0$ V.  
}
\label{figS7}
\end{figure}

We also experimentally demonstrate this nonreciprocal supercurrent effect in our SQUID by applying different values, $V_{g1}$ and $V_{g2}$, to J1 and J2, respectively, as shown in Fig.~\ref{figS7}(b) with $V_{g1} = -4 \text{V}$ and $V_{g2} = 0 \text{V}$. The critical currents of J1 and J2, which are gate-tunable Al-InAs JJs, can be controlled by adjusting $V_{g1}$ and $V_{g2}$. In this case, $B_y = 0$, where the JDE is absent. The different values of $V_{g1}$ and $V_{g2}$ lead to asymmetry in the critical currents of the two JJs. This difference in critical currents results in nonreciprocal critical currents in the SQUID.

\bigskip
\subsection{Nonreciprocity due to JDE}\label{subsec3}

\begin{figure}[h]%
\centering
\includegraphics[width=1\textwidth]{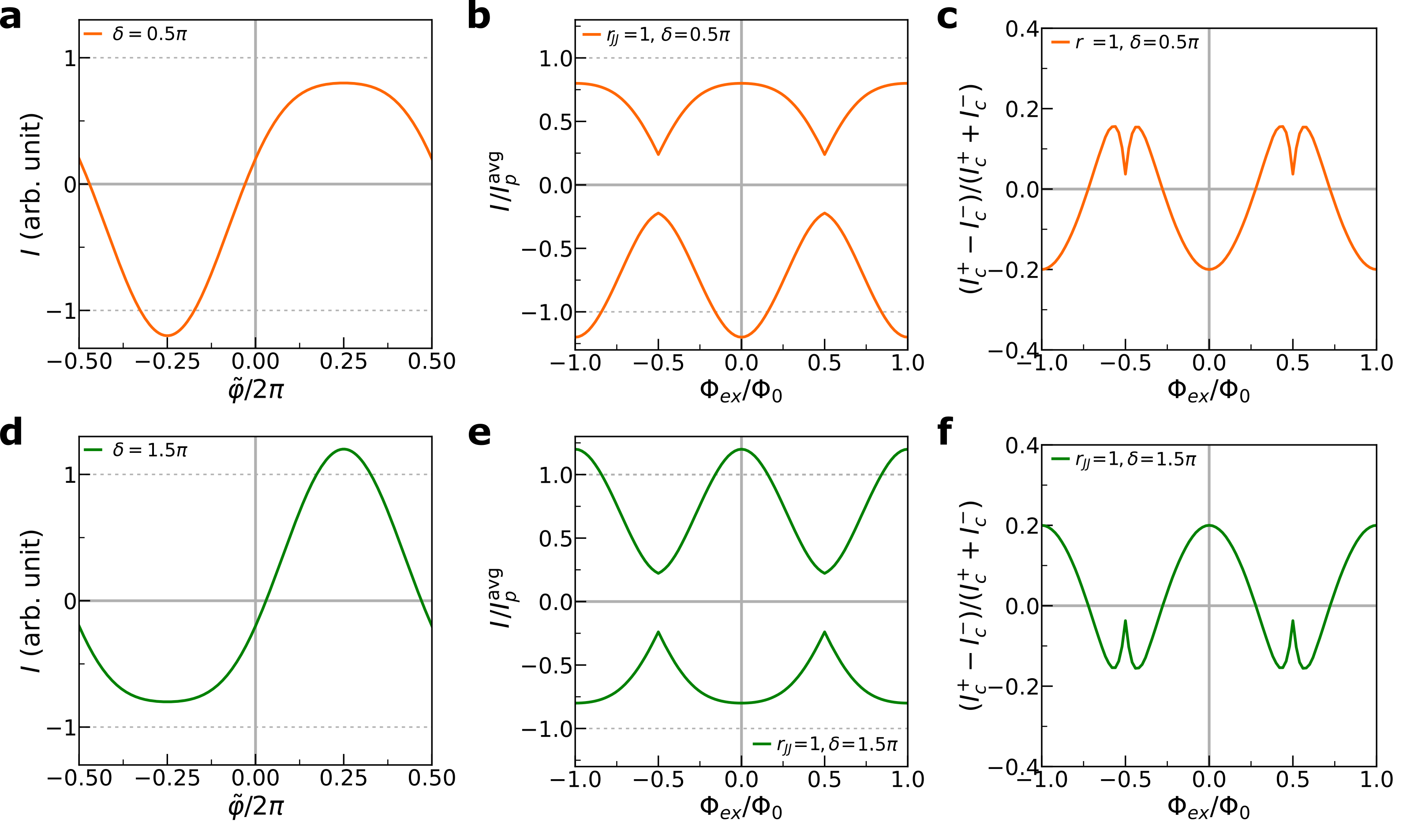}
\caption{
Critical current oscillation of a SQUID with JDE.
(a) CPR in a JJ with $a_{1} = 1$, $a_{2}/a_{1} = 0.2$, and $\delta = \pi/2$, which corresponds to the orange solid line in Fig.~1(c) of the main text.
(b) Maximum and minimum supercurrent in a SQUID, with $a_{1} = 1$, $a_{2}/a_{1} = 0.2$, $\delta = \pi/2$, and $r_{JJ} = 1$, which corresponds to the orange solid line in Fig.~1(d) of the main text.
(c) $(I_{c}^{+} - I_{c}^{-})/(I_{c}^{+} + I_{c}^{-})$ as a function of $\Phi_{ex}/\Phi_{0}$, for the case with $a_{1} = 1$, $a_{2}/a_{1} = 0.2$, $\delta = \pi/2$, and $r_{JJ} = 1$. 
(d) CPR in a JJ with $a_{1} = 1$, $a_{2}/a_{1} = 0.2$, and $\delta = 3\pi/2$.
(e) Maximum and minimum supercurrent in a SQUID, with $a_{1} = 1$, $a_{2}/a_{1} = 0.2$, $\delta = 3\pi/2$, and $r_{JJ} = 1$.
(f) $(I_{c}^{+} - I_{c}^{-})/(I_{c}^{+} + I_{c}^{-})$ as a function of $\Phi_{ex}/\Phi_{0}$, for the case with $a_{1} = 1$, $a_{2}/a_{1} = 0.2$, $\delta = 3\pi/2$, and $r_{JJ} = 1$.  
}
\label{figS8}
\end{figure}

\noindent We demonstrate the distinct nonreciprocal supercurrent effect in a SQUID, which results from the JDE. To induce the JDE in the CPRs of JJs, we set $\delta = \pi/2$ and $3\pi/2$ (Figs.~\ref{figS8}(a) and \ref{figS8}(d)). The other parameters, including $r_{JJ}$, are the same as those in the SQUID oscillation case shown in Fig.~\ref{figS5}(c). For $\delta = \pi/2 $ ($0 < \delta < \pi$), the JDE occurs in the backward current direction, whereas for $\delta = 3\pi/2 $ ($\pi < \delta < 2\pi$), it occurs in the forward current direction. The JDE causes the amplitude of the SQUID oscillations to vary with the direction of the current (Figs.~\ref{figS8}(b) and \ref{figS8}(e)). Although the SQUID oscillation remains symmetric about the $\Phi_{ex}$-axis, it becomes asymmetric about the $I$-axis due to the direction-dependent oscillation amplitude. This leads to a nonreciprocal critical current, maximized at the peaks of the SQUID oscillation, $I_{p}^{+}$ and $I_{p}^{-}$. 

\begin{figure}[h]%
\centering
\includegraphics[width=1\textwidth]{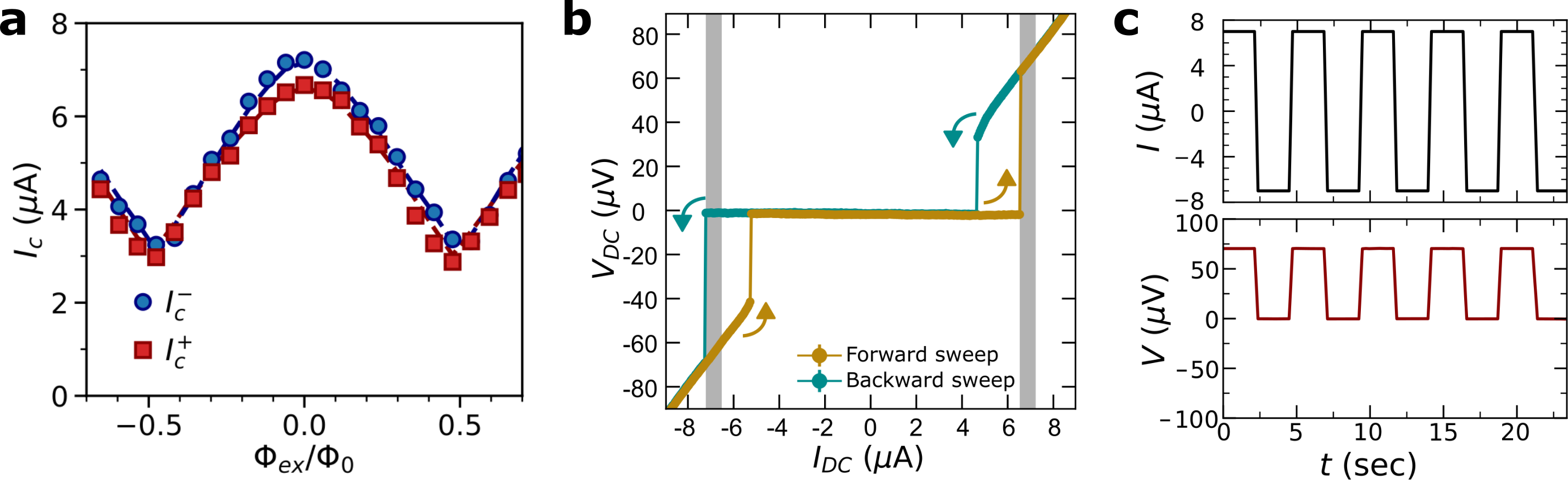}
\caption{
Demonstration of supercurrent rectification due to the JDE.
(a) SQUID oscillations of forward and backward critical currents with $B_{y} = 33$ mT and $V_{g1} = V_{g2} = 0$.
(b) Current-voltage($I$--$V$) characteristics for forward and backward current sweeps at $B_{y} = 33$ mT,  $V_{g1} = V_{g2} = 0$, and $\Phi_{ex}/\Phi_{0} = 0$. The gray-shaded region represents the current range between $I_{c}^{+}$ and $I_{c}^{-}$. 
(c) Supercurrent rectification demonstrated under the same condition as the $I$--$V$ characteristics in (b). The applied current amplitude is 7 $\upmu$A, which falls within the gray-shaded range.
}
\label{figS9}
\end{figure}

Figure~\ref{figS9} demonstrates the supercurrent rectification resulting from the JDE. At $B_{y} = 33 \text{ mT}$ $\Phi_{ex}/\Phi_{0}=0$, and $V_{g1} = V_{g2} = 0 \text{ V}$, $I_{p}^{+}$ is 6.6 $\upmu$A, while $I_{p}^{-}$ is 7.25 $\upmu$A (Fig.~\ref{figS9}(b)). Due to the nonreciprocal critical currents, when a current with an amplitude of 7 $\upmu$A is applied, a dissipationless current is obverved only in the backward direction, whereas a dissipative current flows in the opposite direction (Fig.~\ref{figS9}(c)).\\

\newpage

\newpage

\section{Extraction of anomalous phase difference $\delta$}\label{fit}
We extract the phase difference $\delta$ by fitting SQUID oscillations with numerically calculated SQUID oscillations. The fitting process is performed using a Python script, as shown below.

\bigskip
\lstset{texcl=true,basicstyle=\footnotesize\sf,commentstyle=\small\rm,mathescape=true,escapeinside={(*}{*)}}
\begin{lstlisting}
# In[1]: List of libaries 
import numpy as np
from scipy.optimize import curve_fit
import scipy.constants as const
import math

# In[2]: Define Functions
pi = math.pi
e= const.e
h=const.h
QF = h/(2*e)

#Current phase relation
def CPR_1st(d0,I1,ph1): 
    d = d0+ph1
    return I1*np.sin(d)    
def CPR_2nd(d0,I1,r_2nd,delta_pi,ph1): 
    d = d0+ph1
    delta = delta_pi+pi # delta_pi = delta-pi
    return I1*(r_2nd*np.sin(2*d+delta))
def CPR(d0,I1,r_2nd,delta_pi,ph1):
    return CPR_1st(d0,I1,ph1) + CPR_2nd(d0,I1,r_2nd,delta_pi,ph1)

#Eq.S4
def JJ_e_1(d01,d02,I1,r_2nd,delta_pi,ph1,ph2,r_JJ):
    i = CPR(d01,I1,r_2nd,delta_pi,ph1)+r_JJ*CPR(d02,I1,r_2nd,delta_pi,ph2)  
    return i

#Eq.S5
def JJ_e_2(d01,d02,I1,r_2nd,delta_pi,ph1,ph2,r_JJ):
    j = 0.5*(r_JJ*CPR(d02,I1,r_2nd,delta_pi,ph2)-CPR(d01,I1,r_2nd,delta_pi,ph1))
    return j

#Eq.S6
def function(phi_ex,d01,d02,beta_L,Ic1,r_2nd,delta_pi,ph1,ph2,r_JJ):
    temp = d02 - d01 + (2*pi/QF)*(phi_ex+L*JJ_e_2(d01,d02,Ic1,r_2nd,delta_pi,
    ph1,ph2,r_JJ))
    return temp

def SQUID(phi_a,Ic1,r_2nd,r_JJ,delta_pi,ph2):
    
    d02 = 2*pi*np.linspace(0, 1, num = 501,endpoint = True) 
    d01 = 2*pi*np.linspace(-4, 2, num = 3001,endpoint = True)
    
    Ic = np.zeros(len(phi_a)) 

    L = 0.18E-9
    
    ph1 = 2*pi*0 
    
    for k in range(len(phi_a)):
        phi_ex = phi_a[k]
        i = np.zeros(len(d02))
        for n in range(len(d02)):
            d2 = d02[n]
            temp = function(phi_ex,d01,d2,beta_L,Ic1,r_2nd,delta_pi,ph1,ph2,
            r_JJ) 
            d1 = d01[np.argmin(np.absolute(temp))] 
            i[n] = JJ_e_1(d1,d2,Ic1,r_2nd,delta_pi,ph1,ph2,r_JJ)
        
        # a multi-valued function -> a single-valued function
        if phi_a[k] > 1:   
            Ic[k] = np.amin(i);
        else:
            Ic[k] = np.amax(i);
    return Ic

# In[3]: Import data
data_p = np.loadtxt('data/Bz_Ic_positive_avg_std.txt')
data_n = np.loadtxt('data/Bz_Ic_negative_avg_std.txt')

# In[3]: Fit data
X_temp_p = data_p[3:30,0]/4.2E-5 
X_temp_n = data_n[3:30,0]/4.2E-5

# a multi-valued function -> a single-valued function
X = np.append(X_temp_p,X_temp_n+2) 
Y = np.append((data_p[3:30,1])*1E6,(data_n[3:30,1])*1E6) 

p_i = np.array([3.5,0.1,1,0,0])
popt, pcov = curve_fit(SQUID, X, Y, p0 = p_i, bounds=([0.1,0,0.5,-1*pi,-1*pi],
[5,0.3,2,pi,pi]))
\end{lstlisting}

\begin{figure}[h]%
\centering
\includegraphics[width=1\textwidth]{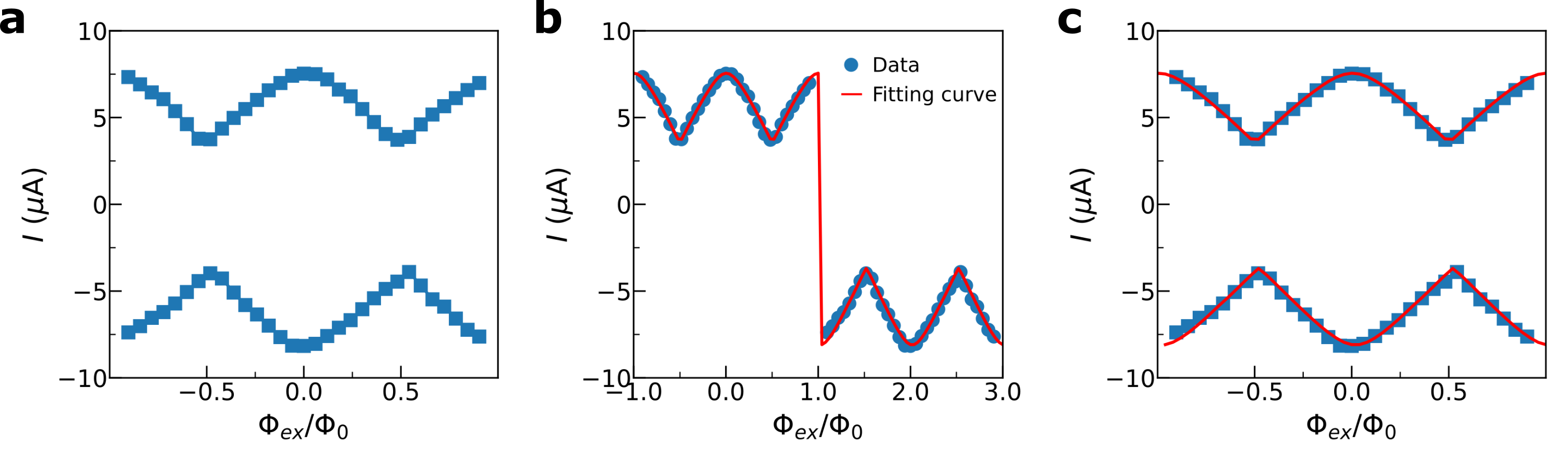}
\caption{
Fitting procedure. 
(a) The data of the SQUID oscillation of forward and backward critical current. 
(b) The fitting result with the modified single-valued SQUID oscillation function. 
(c) The final results of the fitting procedure.
}
\label{figS10}
\end{figure}

The critical current in a SQUID is a multi-valued function of $\Phi_{ex}/\Phi_{0}$ with two values depending on the polarity, $I_{c}^{+}$ and $I_{c}^{-}$. To fit a multi-valued function, we use a simple technique. We modify the multi-valued function to a single-valued function by shifting the backward critical currents along the $\Phi_{ex}/\Phi_{0}$-axis by $+2$ (Fig.~\ref{figS10}(b)). This modification allows us to fit the multi-valued data and extract the second harmonic parameters in the CPR, particularly $\delta$. When presenting the fit result, the shifted backward critical currents and the fit curve are shifted along the $\Phi_{ex}/\Phi_{0}$-axis by $-2$, restoring the modified data to its original form. 

\newpage
\section{Theoretical model}
Given that the two planar JJs forming the SQUID device are identical, we examine a single planar JJ and calculate the Andreev level spectrum and supercurrent by solving the Bogoliubov-de Gennes (BdG) equation for the junction, $H_{\text{BdG}} \Psi = E \Psi$, in the Nambu basis $\Psi = \left(\psi_{e\uparrow}, \psi_{e\downarrow},\psi_{h\downarrow},-\psi_{h\uparrow}\right)^{T}$ with excitation energy $E$ \cite{Gennes1989}. The BdG Hamiltonian $H_{\text{BdG}}$ reads
\begin{align}
  H_{\text{BdG}}&=
  \begin{pmatrix}
      H_0 - \mu & \Delta(x)\\
      \Delta^*(x) & -\mathcal{T} H_0 \mathcal{T}^{-1} + \mu
  \end{pmatrix}, \label{BdG}
\end{align}
where $\mu$ is the chemical potential measured from the bottom of the electron band and $\mathcal{T}=-i \sigma_y \mathcal{C}$ is the time-reversal operator with Pauli matrices $\sigma_i$ ($i=x,y,z$) in spin space and complex conjugation $\mathcal C$. The electron gas with effective mass $m^{*}=0.023\, m_e$ is described by the electron Hamiltonian $H_0$, 
\begin{equation}
 H_0 = \frac{\hbar^2 (k^2_x + k^2_y)}{2 m^*}  -(\alpha +\beta) k_x {\sigma}_y +(\alpha-\beta) k_y {\sigma}_x  + (V_b + E_Z \, \hat{\sigma}_y) L_j \, \delta(x).  \label{h0} 
\end{equation}
It takes into account the Rashba $\alpha$ and Dresselhaus $\beta$ SOCs along the crystallographic [110] direction \cite{Baumgartner_2022, Scharf2019}, the Zeeman energy $E_Z = g \mu_B B_y/2$ of the electrons in the in-plane magnetic field applied along the $y$-direction with $g$-factor $g=-17$ \cite{lee2019}, the potential scattering $V_b$. 
We assume that $E_Z$ and $V_b$ are present only at the junction region $0< x <L_j$ and zero elsewhere. The width of the junction $W = 4$ $\upmu$m  is much larger than $L_j = 100$ nm. The proximity-induced superconducting pairing potential $\Delta(x)$ is given by \cite{davydova2022}
\begin{equation}
  \Delta(x) = \Delta\, e^{i 2 q x} \left[ \Theta(-x)+ e^{i \varphi}\Theta(x-L_j)\right],  
\end{equation}
where $q$ is the orbital induced fCPM, $\Delta = 170 \, \upmu$eV is the proximity-induced gap and $\varphi$ is the superconducting phase difference, and $\Theta(x)$ is the step function.  

We consider a short junction limit $L_j \ll \xi = \hbar v_F/\Delta$. For $\mu=17 
$ meV, which is used in Figure~4, the superconducting coherence length is $\xi= 2$ $\upmu$m. The scattering in the junction is modeled by a delta function $\delta(x)$, as shown in eq~\ref{h0}. We focus on the regime where the chemical potential is much larger than the superconducting gap, $\mu \gg \Delta$, allowing us to neglect the normal reflection at the interface between the superconductor and the normal region can be neglected (Andreev approximation).
Our model in eq~\ref{BdG} then can be treated by linearizing the energy dispersion around $\mu$. We impose hard-wall boundary conditions at $y=0$ and $W$, which quantize the wave vector $k_y$ as $k_m = m \pi/W$. The physical confinement along the $y$-direction results in multiple transverse subbands labeled by $m$. The choice of boundary conditions would be irrelevant in our case of $W\gg L_j$. The total Josephson current flowing across the junction in the $x$-direction is obtained by summing up the contributions from each transverse subband, 
\begin{equation}\label{Current}
  I(\varphi) = -\frac{e}{\hbar}\sum^{M}_{m=1}  \int^{\infty}_{0} dE\, E\, \frac{\partial}{\partial \varphi} \rho_m (E,\varphi),
\end{equation}
where $e>0$ is the elementary charge and $M$ is the number of transverse subbands below the chemical potential. The density of states $\rho_m$ of the junction can be expressed in terms of the scattering matrix $s_N^{(m)}$ of the normal region and the matrix $s_A^{(m)}$ at the superconductor-normal interface as 
\begin{equation}
   \rho_m (E,\varphi)= -\frac{1}{\pi} \text{Im} \frac{\partial}{\partial E} \,\text{Log  Det}\left[I-s^{(m)}_A (E+i \varepsilon,\varphi)\, s^{(m)}_N(E+i \varepsilon,\varphi) \right]. 
\end{equation}
Here we introduced an infinitesimal imaginary energy $\varepsilon$ to calculate the density of states of both the bound and continuum states of the junction. The scattering matrices $s^{(m)}_A $ and $s^{(m)}_N $ are obtained by linearizing the energy dispersion of a transverse subband for a given wave vector $k_m$.

\newpage
\section{Spin-orbit coupling and finite Cooper-pair momentum}\label{sec5}
Our theoretical model accounts for Rashba ($\alpha$) and Dresselhaus ($\beta$) SOCs, the former results from the structural inversion asymmetry along the confinement direction, which is the $z$-direction in our setup, and the latter originates from the lack of inversion symmetry of the crystal structure. In our InAs/InGaAs two-dimensional electron system, the strength of $\alpha$ is tunable and can be controlled by applying an external gate voltage, but Dresselhaus coupling $\beta$ is inherent to the crystal structure and depends on crystallographic direction. Here, we adopt the spin-orbit Hamiltonian from Ref.~\cite{Baumgartner_2022} which corresponds to the current flow along the [110] direction, and $\alpha$ and $\beta$ are considered as fitting parameters. 

\begin{figure}[h]%
\centering
\includegraphics[width=0.8\textwidth]{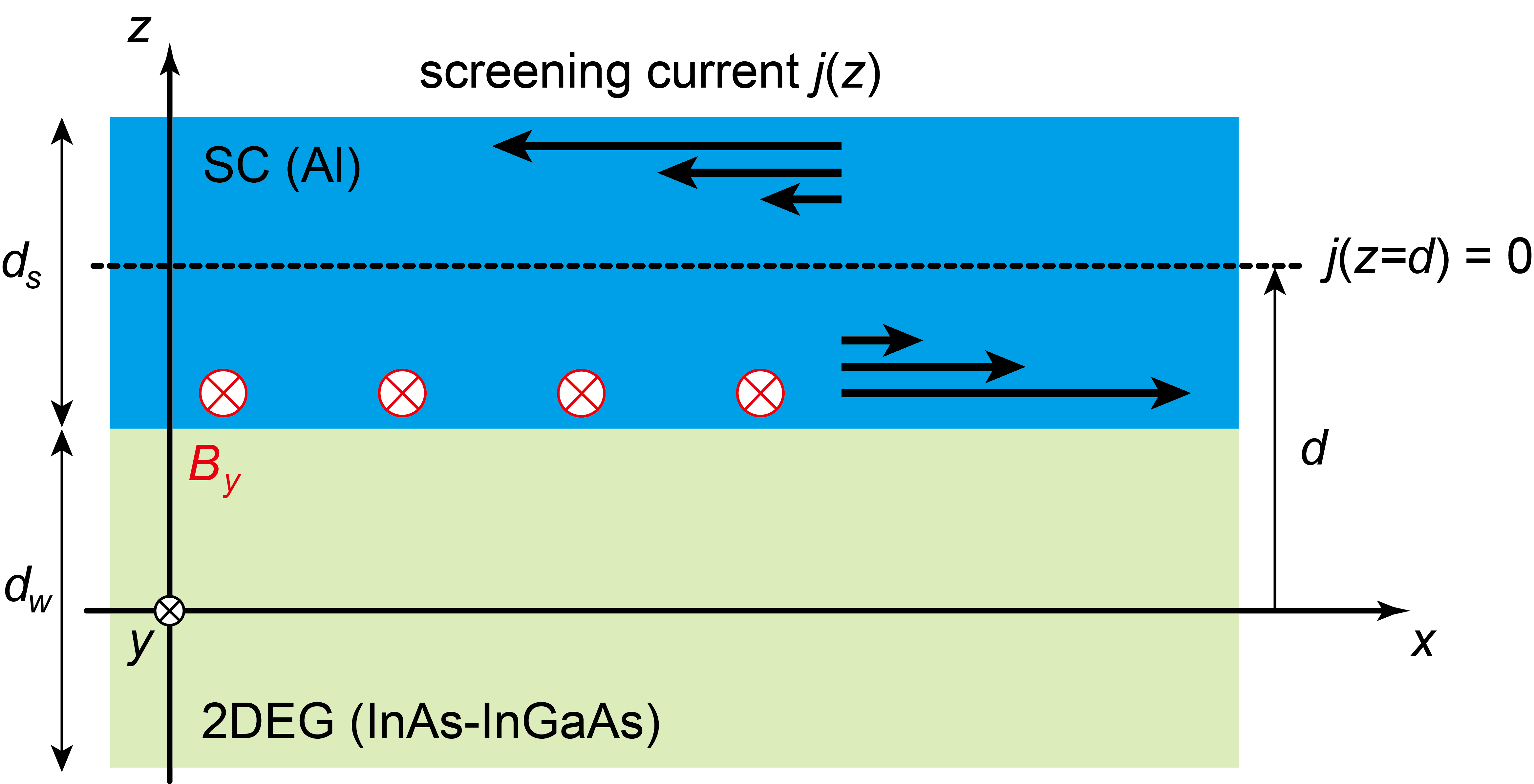}
\caption{
Schematic figure showing orbital induced fCPM. External magnetic field $B_y$ is screened within the bulk SC by generating a screening current near the surfaces. The screening current induces finite momentum to the Cooper pairs in two-dimensional electron gas (2DEG) by proximity effects.
}
\label{figS11}
\end{figure}

The finite momentum $q$ carried by the Cooper pairs due to the orbital effect of the in-plane magnetic field can be obtained from the second Ginzburg-Landau equation of the screening current density~\cite{Clem2010,banerjee2023}, 
\begin{equation}
  \vec{j}= -\frac{1}{\mu_0 \lambda^2} \left( \vec{A}+ \frac{\Phi_0}{2\pi} \nabla \varphi \right),   
\end{equation}
with the permeability $\mu_0$, the London penetration depth $\lambda$, and the superconducting flux quantum $\Phi_0 = h/2e$. The vector potential $ \vec{A} = B_y z \,\hat{x}$ of the magnetic field is set to be zero within the two-dimensional electron gas at $z=0$. 
In our Al-InAs heterostructure illustrated in Fig.~\ref{figS11}, the superconductor is located at $z=d$ above the electron gas where $d=(d_w + d_s)/2$ is one-half of the total thickness of the semiconducting quantum well ($d_w$) and the superconductor ($d_s$). Within the superconducting region $d-d_s/2 < z < d + d_s/2$, we assume that the screening current is uniform in the $y$-direction and it flows in the $x$-direction in opposite direction on the different sides with respect to $z=d$ at which the current is zero $j_x = 0$ if the self-field generated by the screening current is assumed to be negligible. Using the vector potential approximately given by $\vec{A} = B_y d \,\hat{x}$ at $z=d$ due to the thin film geometry, we obtain the phase of the superconducting order parameter,
\begin{equation}
 \varphi = 2 q x + \text{const.}, \,\,  q=-\frac{\pi B_y d}{\Phi_0},  
\end{equation}
where the const. is the integration constant. Without loss of generality, we absorb the constant into $\varphi_0$. Note that the sign of $q$ is reversed if we switch the coordinate frame from right-handed to left-handed~\cite{banerjee2023}.\\

In general, fCPM $q_{\rm soc}$ may be induced by strong Rashba SOC with the magnetic field. In experiments, one can distinguish this spin-orbit-induced Cooper pair momentum from that due to the orbital effect by measuring the gate-voltage dependence of the supercurrent diode effect as the strength of the Rashba coupling is varied by the gate voltage. In our experiments, the gate-voltage dependence is very weak at low magnetic field, indicating that the Cooper pair momentum is due to the orbital effect. 
With the parameters of $\alpha = 10$ meV nm and $v_F = 5.1\times 10 ^5$ m/s used in our theoretical calculation, we estimate the value of $q_{\rm soc}$~\cite{yuan2022}, 
\begin{equation}
 |q_{\rm soc}| =\Big|\alpha \, \frac{1}{\hbar^2 v^2_F} \frac{g \mu_B B_y}{2}\Big|\approx 4.41\times 10^{-8} B_y \, \text{mT}^{-1} \text{nm}^{-1}, 
\end{equation}
which is three orders of magnitude smaller than the orbital-induced momentum $q \approx 1.5\times 10^{-5} B_y$ mT$^{-1}$nm$^{-1}$. Therefore, we neglected this effect in the calculation.

\newpage
\section{Scattering matrices of the Josephson junction}
In this section, we provide the details of the calculation of the CPR through scattering formalism in a single JJ. The Bogoliubov-de Gennes (BdG) equation for the JJ is
\begin{eqnarray}
    H_{\rm BdG} \Psi = E \Psi, \ \ \ 
    H_{\rm BdG} = \begin{pmatrix}
        H_0 - \mu & \Delta(x) \\
        \Delta^*(x) & -\mathcal T H_0 \mathcal T^{-1} + \mu
    \end{pmatrix},
\label{eq:BdG}
\end{eqnarray}
where $\Psi = (\psi_{e\uparrow}, \psi_{e\downarrow}, \psi_{h \downarrow}, -\psi_{h \uparrow})^T$ is the Nambu spinor, $E$ is the excitation energy, $H_0$ is the normal state Hamiltonian, $\mu$ is the chemical potential measured from the bottom of the electron band. $\mathcal T = -i\sigma_y \mathcal C$ is the time-reversal operator with Pauli matrices $\sigma_i$ ($i=x,y,z$) and complex conjugation $\mathcal C$, and $\Delta(x)$ is the pairing potential varying along the junction direction $x$. The form of $\Delta(x)$ is given by
\begin{eqnarray}
    \Delta(x) = \Delta e^{i2qx} \big[ \Theta(-x) + e^{i\varphi} \Theta(x-L_j) \big],
\end{eqnarray}
where $q$ is orbital-induced fCPM discussed in Sec.~\ref{sec5}, $\Delta = 170$ $\upmu$eV is the proximity-induced gap, $\varphi$ is the superconducting phase difference, $L_j$ is the length of the junction, and $\Theta(x)$ is the step function. For the normal state Hamiltonian $H_0$, we use a two-dimensional electron gas model with SOC under the applied magnetic field along the $y$-direction:
\begin{eqnarray}
    H_0(k_x, k_y) = \frac{\hbar^2 (k_x^2 + k_y^2)}{2m^*} - (\alpha + \beta) k_x \sigma_y + (\alpha - \beta) k_y \sigma_x + (V_b + E_Z \sigma_y) L_j \delta(x).
\label{eq:H0}
\end{eqnarray}
Here, $m^* = 0.023 m_e$ is the effective mass, $\alpha$ is the Rashba SOC constant, and $\beta$ is the Dresselhaus SOC constant along [110] crystallographic direction. $V_b$ represents potential scattering strength, $E_Z = g\mu_B B_y/2$ is the Zeeman energy with $g$-factor $g=-17$. We assume that $E_Z$ and $V_b$ are present only at the junction region $0< x <L_j$ and zero elsewhere. The width of the junction $W = 4$ $\upmu$m  is much larger than $L_j = 100$ nm.

We consider a short junction limit $L_j \ll \xi = \hbar v_F/\Delta$. For $\mu=17$ meV, the superconducting coherence length is $\xi= 2$ $\upmu$m, and the short junction condition is well satisfied. Therefore, we approximate the scattering processes occurring at $x=0$, as shown in the last term in Eq.~\ref{eq:H0}. In addition, we focus on the regime where the chemical potential is much larger than the superconducting gap, $\mu \gg \Delta$, allowing us to neglect the normal reflection at the interface between the superconductor and the normal region can be neglected (Andreev approximation).

The problem can be reduced to quasi-one-dimensional problem with multiple transverse modes. Imposing hard-wall boundary conditions at $y=0$ and $W$ on $\Psi(x)$, the wave vector $k_y$ becomes quantized as $k_m = m \pi/W$, where $m \in \mathbb Z$. The physical confinement along the $y$-direction results in multiple transverse subbands labeled by $m$. The choice of boundary conditions would be irrelevant in our case of $W\gg L_j$. Then, the normal state Hamiltonian for given transverse mode $m$ is,
\begin{eqnarray}
    H_0(k_x;m) = \frac{\hbar^2 (k_x^2 + k_m^2)}{2m^*} - (\alpha + \beta) k_x \sigma_y + (\alpha - \beta) k_m \sigma_x + (V_b + E_Z \sigma_y) L_j \delta(x).
\end{eqnarray}

\begin{figure}[h]%
\centering
\includegraphics[width=0.9\textwidth]{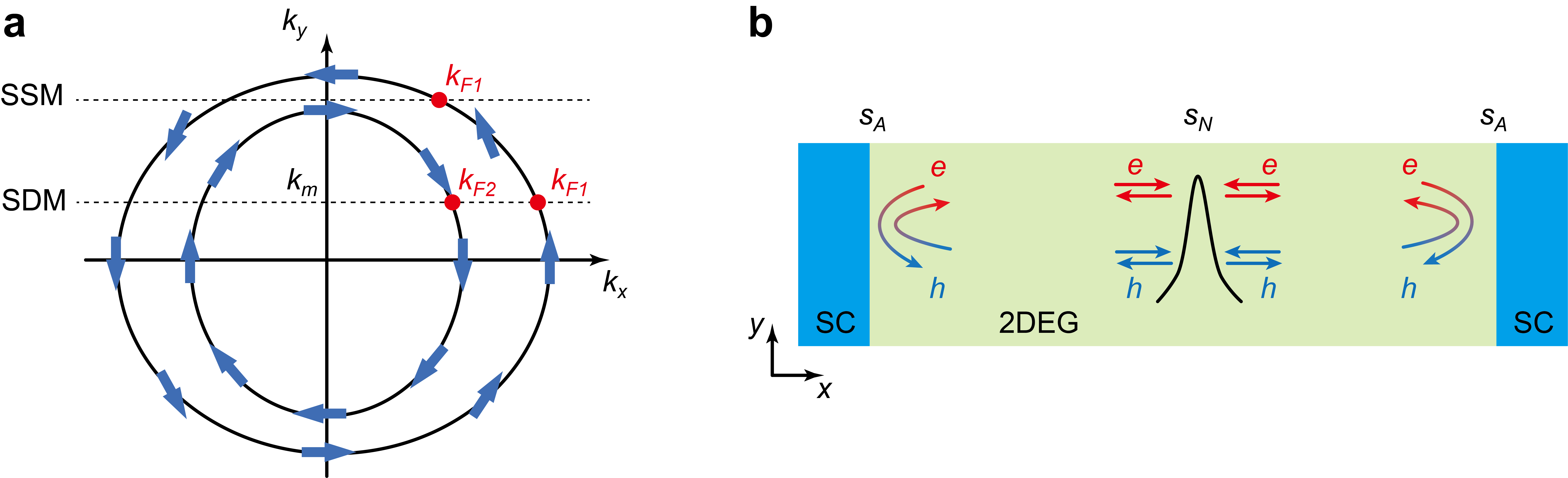}
\caption{
(a) Two spin-split Fermi surfaces due to SOCs. The spin direction (blue arrow) is locked at each Fermi momentum. Spin-split modes (SSM) and spin-degenerate modes (SDM) have one ($k_{F1}$) and two fermi points ($k_{F1}$ and $k_{F2}$), respectively. 
(b) Schematic figure for the scatterings occurring in the Josephson junction. The normal scattering ($s_N$) scatters electrons (holes) to electrons (holes), while the Andreev reflections occurring at the superconducting-normal metal interfaces convert electrons (holes) to holes (electrons).
}
\label{figS12}
\end{figure}

The number of bands of the effective one-dimensional Hamiltonian $H_0(k_x;m)$ can be one or two depending on the value of $m$, originating from the spin-split two-dimensional bands of $H_0(k_x, k_y)$ by the SOCs. We refer to the modes with single-band with $k_{y2}<|k_m|<k_{y1}$ and double-band $|k_m|<k_{y2}$ as spin-split mode (SSM) and spin-degenerate mode (SDM), respectively (Fig.~\ref{figS12}(a)). Here,
\begin{eqnarray}
    k_{y1} = \sqrt{ k^2_{F} + \frac{1}{4} (\tilde\alpha - \tilde\beta)^2} + \frac{|\tilde\alpha - \tilde\beta|}{2}, \ \ \ 
    k_{y2} = \sqrt{ k^2_{F} + \frac{1}{4} (\tilde\alpha - \tilde\beta)^2} -\frac{|\tilde\alpha - \tilde\beta|}{2},
\end{eqnarray}
with $k_{F} = \sqrt{2m^*\mu}/\hbar$, $\tilde\alpha = 2m^*\alpha/\hbar^2$, and $\tilde\beta = 2m^*\beta/\hbar^2$. For convenience, we provide the eigenvalues and eigenstates of $H_0(k_x;m)-\mu$:
\begin{eqnarray}
    \epsilon_\pm(k_x) = \frac{\hbar^2(k_x^2 + k_m^2)}{2m^*} - \mu \pm \sqrt{\alpha_+^2 k_x^2 + \alpha_-^2 k_m^2}, \ \ \ 
   \chi_\pm(k_x) = \frac{1}{\sqrt 2} \begin{pmatrix}
       1\\
        \mp e^{i \theta_{\rm soc}}
    \end{pmatrix}, 
\end{eqnarray}
where $\alpha_\pm = \alpha \pm \beta$ and 
\begin{equation}
   e^{i \theta_{\rm soc}} = \frac{i\alpha_+ k_x - \alpha_- k_m}{\sqrt{\alpha_+^2 k_x^2 + \alpha_-^2 k_m^2}}. \label{soc-angle}
\end{equation}
Below, we provide the scattering matrices $s^{(m)}_N$ and $s^{(m)}_A$ in detail. We will drop the mode index $m$ for convenience unless indicated otherwise.

\newpage
\subsection{Spin-split mode (SSM)}

\noindent There are two Fermi points at $k_x = \pm k_{F_1}$ from which $\epsilon_{-}(k_x)=0$, 
\begin{eqnarray}
    k_{F_1} &=& \sqrt{ k_{F}^2 - k_m^2 + \frac{(\tilde \alpha + \tilde \beta)^2}{2} + \sqrt{(\tilde\alpha+\tilde\beta)^2 k_{F}^2 + \frac{(\tilde\alpha+\tilde\beta)^4}{4} - 4\tilde{\alpha}\tilde{\beta} k_m^2}}.
\end{eqnarray}
Only the modes near the Fermi points are relevant in a low-energy limit. Then, one can write the electron wave function $\psi_e(x)$ as a linear combination of the left- and right-moving states,
\begin{eqnarray}
    \psi_e(x) = e^{ik_{F_1}x} \chi_R \psi_{eR}(x) + e^{-ik_{F_1}x} \chi_L \psi_{eL}(x),
\end{eqnarray}
where $\chi_{R} = \chi_{-}(k_{F_1})$ and $\chi_{L} = \chi_{-}(-k_{F_1})$ are the right-moving and left-moving eigenstates of $H_0(k_x;m)-\mu$ at $k_x=\pm k_{F_1}$, respectively. Neglecting the rapidly oscillating modes ($\sim e^{\pm 2ik_{F_1}x}$), we obtain
\begin{eqnarray}
    \tilde H_0 -\mu = \begin{pmatrix}
        \hbar v_{F_1} (k_x -k_{F_1}) & 0 \\ 0 & -\hbar v_{F_1} (k_x+k_{F_1})
    \end{pmatrix}, \ \ \ v_{F_1} = \frac{1}{\hbar} \frac{\partial \epsilon_-(k_x)}{\partial k_x} \bigg|_{k_x = k_{F_1}}.
\end{eqnarray}
We will neglect the $-\hbar v_F k_{F_1}$ term in the Hamiltonian henceforth for simplicity, as it does not affect the scattering matrices. 
Note that the spin degree of freedom is locked, so $\tilde H_0$ is still 2$\times$2 matrix. The same procedure for the hole part yields the linearized Hamiltonian for the basis $\tilde\Psi = (\psi_{eR}, \psi_{eL}, \psi_{hR}, \psi_{hL})^T$, 
\begin{eqnarray}
    \tilde H_{\rm BdG} \tilde\Psi = E \tilde\Psi, \ \ \
    \tilde H_{\rm BdG} = \begin{pmatrix}
        -i\hbar v_{F_1} \hat d_z \partial_x & \Delta(x) \\
        \Delta^*(x) & i\hbar v_{F_1} \hat d_z \partial_x 
    \end{pmatrix}+ 
    \begin{pmatrix}
      \hat{V} + \hat{Z} & 0 \\
      0 & -\hat{V} + \hat{Z}
    \end{pmatrix} \delta(x),
\end{eqnarray}
where $\hat d_i$ ($i=x,y,z$) is Pauli matrices in the right/left mover space. The pairing potential $\Delta(x)$ is diagonal in the right-left mover space because of the electron-hole symmetry. The barrier terms $\hat{V}$ and $\hat{Z}$ are written as
\begin{eqnarray}
    \hat{V} &=& \begin{pmatrix}
        \bra{\chi_R} V_b L_j \ket{\chi_R} & \bra{\chi_R} V_b L_j \ket{\chi_L} \\
        \bra{\chi_L} V_b L_j \ket{\chi_R} & \bra{\chi_L} V_b L_j \ket{\chi_L}
    \end{pmatrix}, \\
     \hat{Z} &=& \begin{pmatrix}
        \bra{\chi_R} E_z\sigma_y L_j \ket{\chi_R} & \bra{\chi_R} E_z\sigma_y L_j \ket{\chi_L} \\
        \bra{\chi_L} E_z\sigma_y L_j \ket{\chi_R} & \bra{\chi_L} E_z\sigma_y L_j \ket{\chi_L}
    \end{pmatrix}.
\end{eqnarray}
With Pauli matrices $\hat \tau_i$ ($i=x,y,z$) acting on the particle-hole space, we perform a unitary transformation by 
$\hat{U} = \exp (-i\frac{\varphi}{2} \Theta(x) \hat\tau_z - iqx \hat\tau_z)$, resulting in
\begin{eqnarray}
   H_{\rm SSM} \Psi_{\rm SSM} = E \Psi_{\rm SSM}, \,\,\,  H_{\rm SSM}\equiv \hat{U} \tilde H_{\rm BdG} \hat{U}^{\dagger} = H_{kin} + H_b \delta(x) + H_{sc},
\end{eqnarray}
where
\begin{eqnarray}
    H_{kin} &=& -i\hbar v_{F_1} \hat d_z \hat\tau_z \partial_x + \hbar v_{F_1} q \hat d_z, \nonumber \\
    H_{b} &=& \hat{V} \hat{\tau}_z + \hat{Z} + \frac{\hbar v_{F_1} \varphi}{2} \hat d_z, \nonumber \\
    H_{sc} &=& \Delta \big[ \Theta(-x) + \Theta(x-L_j) \big]. \nonumber 
\end{eqnarray}
The boundary condition of the states across the scatterer at $x=0$ is then imposed as 
\begin{equation}
   \Psi_{\rm SSM}(0^{+}) = \exp \left(-\frac{i}{\hbar v_{F_1}} \hat d_z \hat\tau_z H_b \right) \Psi_{\rm SSM}(-0^{+}), 
\end{equation}
where $0^+$ is the positive infinitesimal. We convert this transfer matrix to scattering matrix $s_N$ which connects the incoming and outgoing states with respect to $x=0$ (Fig.~\ref{figS12}(b)),  
\begin{eqnarray}
    \Psi_{\rm out} &=& s_N \Psi_{\rm in}, \\
    \Psi_{\rm in}(x) &=& \begin{pmatrix}
        \psi_{eR}(x) \\ 0 \\ 0 \\ \psi_{hL}(x)
    \end{pmatrix} \Theta(-x)
    + \begin{pmatrix}
        0 \\ \psi_{eL}(x) \\ \psi_{hR}(x) \\ 0
    \end{pmatrix} \Theta(x), \\
    \Psi_{\rm out}(x) &=& \begin{pmatrix}
        0 \\ \psi_{eL}(x) \\ \psi_{hR}(x) \\ 0
    \end{pmatrix} \Theta(-x)
    + \begin{pmatrix}
        \psi_{eR}(x) \\ 0 \\ 0 \\ \psi_{hL}(x)
    \end{pmatrix} \Theta(x).
\end{eqnarray}
The matrix $s_N$ is given by
\begin{eqnarray}
    s_N = \begin{pmatrix}
        s_0 & 0 \\ 0 & s_0^*
    \end{pmatrix}, \ \ \ 
    s_0 = \begin{pmatrix}
        r &  e^{i\frac{\varphi}{2}} t' \\
        e^{-i\frac{\varphi}{2}} t & r'
    \end{pmatrix}.
\label{sn-matrix}
\end{eqnarray}
The reflection and transmission coefficients are determined by 
\begin{align}
 t e^{i \theta_z} &= t' e^{-i \theta_z} = \left(\cos d + i n_z \sin d \right)^{-1}, \nonumber\\
 r e^{-i \theta_{\rm soc}} &=r' e^{i \theta_{\rm soc}} = -i n_z  \sqrt{t t'}\,  \cos \theta_{\rm soc}\, \sin d, \nonumber\\ 
 \theta_z &= \frac{E_z L_j \sin \theta_{\rm soc}}{\hbar v_{F_1}}, \,\,\, d= \left| \frac{V_b L_j\sin \theta_{\rm soc}}{\hbar v_{F_1}} \right|,  
\end{align}
where $n_z = |\sin \theta_{\rm soc}|^{-1}$ and $\theta_{\rm soc}$ is defined in Eq.~\ref{soc-angle}.
The Andreev scattering matrix $s_A$ defined by $\Psi_{\rm in}^A = s_A \Psi_{\rm out}^A $ can also be obtained by matching the states at the SN interfaces $|x|=L_j/2$ (Fig.~\ref{figS12}(b)),  
\begin{eqnarray}
    s_A = \begin{pmatrix}
        0 & s_{eh}\\ s_{he} & 0
    \end{pmatrix}, \ \ \ 
    s_{eh} = \hat{d}_x s_{he} \hat{d}_x =\begin{pmatrix}
        \beta_- & 0 \\ 0 & \beta_+
    \end{pmatrix},  \label{asm}
\end{eqnarray}
with
\begin{eqnarray}
    \beta_\pm = \frac{E \pm \hbar v_{F_1} q}{\Delta} - i\sqrt{1 - \frac{(E \pm \hbar v_{F_1} q)^2}{\Delta^2}}.
\end{eqnarray}

\bigskip
\subsection{Spin-degenerate mode (SDM)}

\noindent For SDMs with double bands, two additional Fermi points at $k_x = \pm k_{F_2}$, where $\epsilon_{+}(k_x)=0$, appear on the inner Fermi surface, 
\begin{eqnarray}
    k_{F_2} = \sqrt{ k_{F}^2 - k_m^2 + \frac{(\tilde \alpha + \tilde \beta)^2}{2} - \sqrt{(\tilde\alpha+\tilde\beta)^2 k_{F}^2 + \frac{(\tilde\alpha+\tilde\beta)^4}{4} - 4\tilde{\alpha}\tilde{\beta} k_m^2}}.
\end{eqnarray}
Following the same procedure as for SSMs, we linearize the Hamiltonian and perform the unitary transformation. 
In the basis $\Psi_{\rm SDM} = (\psi_{eR_1}, \psi_{eR_2}, \psi_{eL_1}, \psi_{eL_2}, \psi_{hR_1}, \psi_{hR_2}, \psi_{hL_1}, \psi_{hL_2})^T$, the 8$\times$8 scattering matrices $s_N$ and $s_A$ are obtained. The scattering matrix of electron, $s_0$ in Eq.~\ref{sn-matrix}, has the form 
\begin{equation}
  s_0 = \begin{pmatrix}
  \hat{r} & e^{i \frac{\varphi}{2}} \hat{t}' \\
  e^{-i \frac{\varphi}{2}} \hat{t} & \hat{r}'
\end{pmatrix},
\end{equation}
where $\hat{r}$ and $\hat{r}'$ ($\hat{t}$ and $\hat{t}'$) are the 2$\times$2 reflection (transmission) matrices describing the scattering between the outer band and the inner band.
The matrices $s_{eh}$ and $s_{he}$ of the Andreev scattering matrix $s_A$ in Eq.~\ref{asm} are obtained by   
\begin{equation}
   s_{eh} = \hat{d}_x s_{he} \hat{d}_x =\begin{pmatrix}
        \hat{\beta}_- & 0 \\ 0 & \hat{\beta}_+
    \end{pmatrix}, \,\,\,
  \hat{\beta}_{\mp} = 
  \begin{pmatrix}
   \beta_{\mp 1} & 0 \\
   0 & \beta_{\mp 2}
  \end{pmatrix},
\end{equation}
where $\beta_{\mp j}$ with $j=1,2$ are given by 
\begin{equation}
    \beta_{\mp j} = \frac{E\mp\hbar v_{F_j} q}{\Delta}-i \sqrt{1-\left( \frac{E\mp\hbar v_{F_j} q}{\Delta}\right)^2}.
\end{equation}

\newpage
\section{Evolution of the higher harmonics with magnetic field}

\begin{figure}[h!]%
\centering
\includegraphics[width=0.94\textwidth]{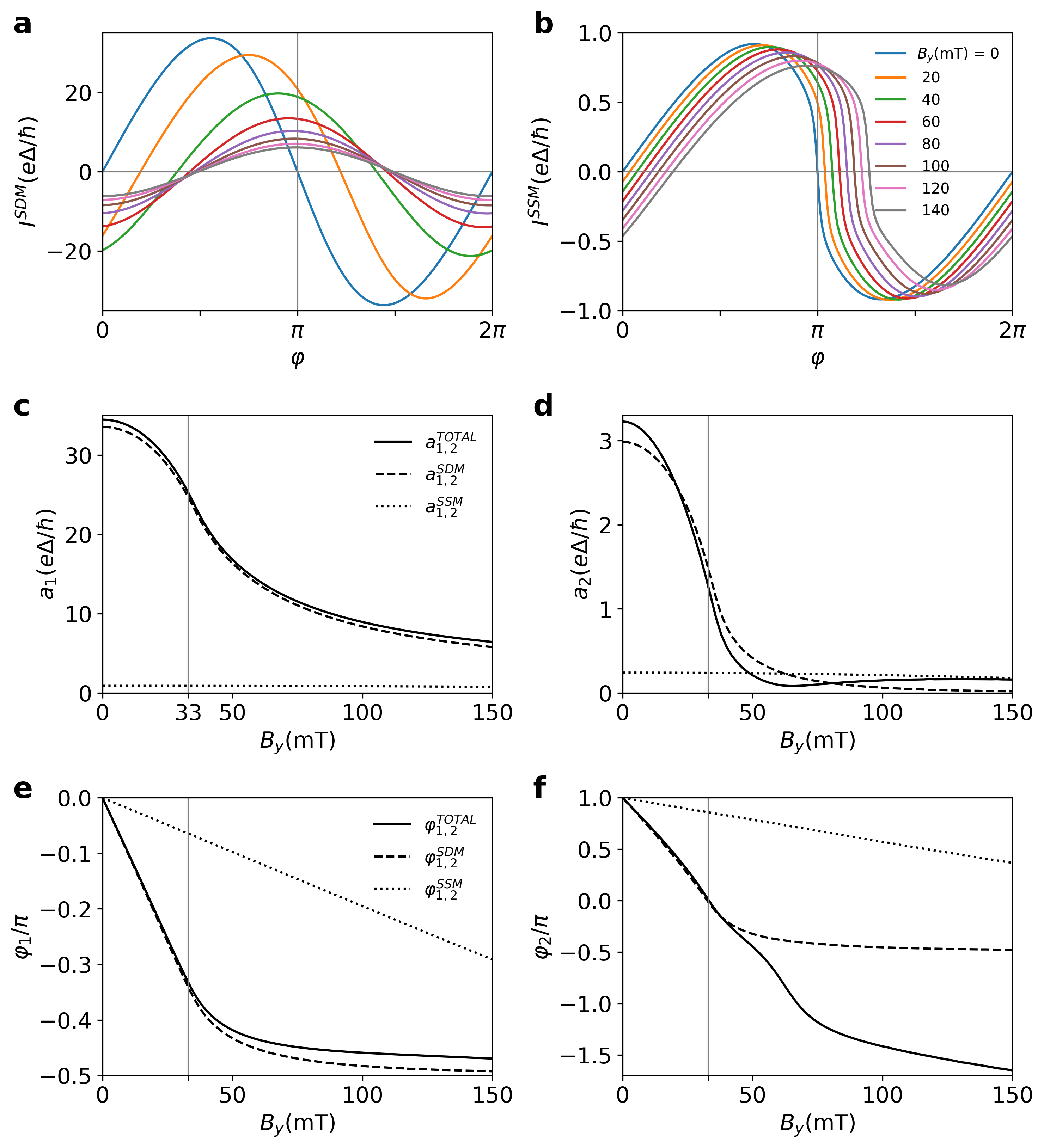}
\caption{
Evolution of the CPR with $B_y$.
(a) The evolution of the CPR for SDM. 
(b) The same as (a), but for SSM.
(c) Amplitudes of the first harmonic obtained by fitting the CPRs in (a) and (b) to Eq. (1) in the main text. 
(d) Amplitudes of the second harmonic obtained as in (c). 
(e) Anomalous phase shift of the first harmonic obtained as in (c). 
(f) Anomalous phase shift of the second harmonic obtained as in (c). 
}
\label{figS13}
\end{figure}

We analyze the evolution of the CPR under the magnetic field $B_y$. 
The CPRs we discuss here correspond to the case of $q\neq 0$ and SOC $\neq$ 0 presented in Fig.~4 in the main text. 
The results are shown in Fig.~\ref{figS13}. 

The plots (a) and (b) in Fig.~\ref{figS13} illustrate how the CPRs evolve with the magnetic field $B_y$. The CPRs for SDM and SSM show distinct characteristics, such that while $I^{\rm SDM}(\varphi)$ exhibits the evolution similar to that driven by the fCPM without SOC~\cite{davydova2022}, $I^{\rm SSM}(\varphi)$ is shifted with $B_y$, reflecting the Zeeman effect on the spin-split states. The plots (c) and (d) provide further analysis by showing the amplitudes of the first and the second harmonics obtained from fitting the CPRs with the model in Eq. (1) in the main text. Note that the amplitudes of the second harmonics, $a^{\rm SDM}_2$ and $a^{\rm SSM}_2$, in (d) are comparable, demonstrating a strong dependence of the anomalous phase shift of the second harmonic on $B_y$, as shown in (f). Such dependence can be seen the following relations, 
\begin{align}
  &a^{\rm SDM}_2 \sin \left(2 \varphi+\varphi^{\rm SDM}_2\right) 
  + a^{\rm SSM}_2 \sin \left(2 \varphi+\varphi^{\rm SSM}_2\right) = a^{\rm TOTAL}_2 \sin\left(2 \varphi +\varphi^{\rm TOTAL}_2\right), \nonumber\\
  &\varphi^{\rm TOTAL}_2 = \arctan\left(\frac{a^{\rm SDM}_2 \sin \varphi^{\rm SDM}_2+a^{\rm SSM}_2 \sin \varphi^{\rm SSM}_2 }{a^{\rm SDM}_2 \cos \varphi^{\rm SDM}_2+a^{\rm SSM}_2 \cos \varphi^{\rm SSM}_2} \right).
\end{align}
Therefore, this pronounced variation is attributed to the comparable amplitudes shown in (d). 

\newpage

\section{Calculation parameters for Figures~4 and 5 in the main text}\label{fit parameters}
Theoretical results were obtained from eq~\ref{Current}. We give in Table~\ref{tab:parameters} the parameters corresponding to the figures. Here, $\Delta = 170$ $\upmu$eV and $q= 1.42 \times 10^{-5} B_y $ mT$^{-1}$nm$^{-1}$ are used. The transparency $\tau$ at zero field is obtained by fitting the calculated CPR with the formula of the form $I_0\, \sin(\varphi)/ \sqrt{1-\tau\, \sin^2(\varphi/2)}$ \cite{Beenakker1991}.

\begin{table}[h!]
    \centering
    \begin{tabular}{|c|c|c|c|c|c|}
    \hline
    ~ & $\alpha$(meV nm)& $\beta$(meV nm) & $\mu$(meV) & $V_b$(meV) & $\tau(B_y=0)$  \\ \hline
    Figures~4a,b, SOC $\neq$ 0 &  7.53 & 4.23 & 17 & 2.76 & 0.536 \\ \hline
    Figures.~4a,b, SOC $=$ 0 &  0 & 0 & 17 & 2.76 & 0.515 \\ \hline
    Figures~5a,b, $V_g$ = $0$ V &  7.53 & 4.23 & 17 & 2.76 & 0.536\\ \hline
    Figures~5a,b, $V_g$ = $-3$ V &  7.26 & 4.23 & 13.6 & 2.51 & 0.529 \\ \hline
    Figures~5a,b, $V_g$ = $-5$ V &  6.03 & 4.23 & 8.5 & 1.84 & 0.542 \\ \hline
    Figures~5a,b, $V_g$ = $-6$ V &  5.08 & 4.23 & 6.8 & 1.34 & 0.636  \\ \hline
    \end{tabular}
    \caption{Parameters used for the calculations of the diode efficiency and the anomalous phase difference shown in the figures, using eq~\ref{Current}.}
    \label{tab:parameters}
\end{table}

\newpage
\RaggedRight
\bibliography{JDE}
\bibliographystyle{JDE}